\pdfoutput=1
\documentclass[useAMS]{mn2e}
\usepackage{times}
\usepackage{amssymb}
\usepackage{amsmath}
\usepackage{rotating}
\usepackage{color}
\usepackage{epsfig}
\usepackage{graphicx}
\input{epsf}

\title[Complex NLS1s: high spin or high inclination?]
{Complex Narrow Line Seyfert 1s: High spin or high inclination?}

\author[E. Gardner and C. Done]
{Emma Gardner and Chris Done\\
Department of Physics, University of Durham, South Road,
Durham DH1 3LE, UK\\}

\date{Submitted to MNRAS}
\pagerange{\pageref{firstpage}--\pageref{lastpage}} \pubyear{}

\begin{document}

\topmargin = -0.5cm

\maketitle

\label{firstpage}

\begin{abstract}

Complex narrow line Seyfert 1s (NLS1s), such as 1H0707-495, differ
from simple NLS1s like PG1244+026 by showing stronger broad spectral
features at Fe K and larger amplitude flux variability. These are
correlated: the strongest Fe K features are seen during deep dips in
the light curves of complex NLS1s. There are two competing
explanations for these features, one where a compact X-ray source on
the spin axis of a highly spinning black hole approaches the horizon
and the consequent strong relativistic effects focus the intrinsic flux
onto the inner edge of a thin disc, giving a dim, reflection dominated
spectrum. The other is that the deep dips are caused by complex
absorption by clumps close to the hard X-ray source. The reflection 
dominated model is able to reproduce the very short 30s soft lag from
reverberation seen in the complex NLS1 1H0707-495. However, it does not explain the
characteristic switch to hard lags on longer timescales. Instead, a
full model of propagating fluctuations coupled to reverberation can
explain the switch in the simple NLS1 PG1244+026 using a low spin
black hole. However PG1244+026 has a longer reverberation lag of $\sim 200$s.

Here we extend the successful propagation-reverberation model for the
simple NLS1 PG1244+026 to include the effect of absorption from clumps in a
turbulent region above the disk. The resulting occultations of the inner accretion flow can introduce
additional hard lags when relativistic
effects are taken into account. This dilutes the soft lag from
reverberation and shifts it to higher frequencies, making a smooth
transition between the 200s lags seen in simple NLS1s to the 30s lags
in complex NLS1s. These two classes of NLS1 could then be determined by
inclination angle with respect to a clumpy, probably turbulent, failed
wind structure on the disc.

\end{abstract}


\section{Introduction} \label{sec:introduction}

Narrow Line Seyfert Type 1s (NLS1s) are small mass, high mass
accretion rate Active Galactic Nuclei (AGN).  Like other AGN, their
fastest, highest amplitude variability is at X-ray energies, but their
lower mass means that this variability is on shorter timescales,
making it easier to study over a typical $<1$ day observation
(e.g. Ponti et al 2010).

All the NLS1s show rapid X-ray variability (Leighly 1999), but some
also show deep dips in the X-ray light curve. These dips coincide with the
appearance of high energy complexity in the 2-10~keV spectra, either
gradual curvature or strong features around the Fe K$\alpha$ line
energy. Gallo (2006) termed these 'complex' NLS1s to distinguish them
from the 'simple' NLS1s which do not show dips and have relatively
power law like spectra from 2-10~keV.  Two different models have been
proposed to explain the deep dips and associated spectral complexity:
partial covering and relativistic reflection.

In the partial covering model, the dips are caused by low ionisation
material moving into the line of sight, increasing the absorption at
the iron edge energy at 7.1~keV. This material can only partially
cover the source as some fraction of the flux at low energies is still
seen (e.g. Inoue et al 2003; Turner et al 2007; Miyakawa et al 2012).
Conversely, in the relativistic reflection model the dips are caused
by an extremely compact X-ray source on the spin axis of the black
hole approaching the event horizon. The resulting strong light bending
focusses the intrinsic continuum away from the observer (producing the
drop in flux) so it instead strongly illuminates the very inner
disc. For high spin black holes the resulting spectrum can be
dominated by highly smeared relativistic reflection, marked by a strong
but extremely broad and skewed Fe K$\alpha$ line (e.g. Fabian et al
2004; Miniutti \& Fabian 2004; Fabian et al 2009). In both models, the
'complex' and 'simple' NLS1s are intrinsically similar, and can change
from one to the other (as observed: Gallo 2006) depending on whether
there is absorption along the line of sight, or in the reflection model, whether the compact X-ray source is close to the
horizon.

Both absorption and reflection models can fit the observed 0.3-10~keV
spectra, as fitting complex models over a limited bandpass is highly
degenerate. However a recent breakthrough is in the use of
variability to break some of these degeneracies. In particular, the
new techniques which can identify a lagged signal are clearly well
matched to the reflection models, where there is a strong prediction
that the reflected emission should lag behind the intrinsic continuum
variability on the light travel time (Fabian et al 2009; Uttley et al
2014).  Detection of a very short lag time ($\sim 30$~s) in the
complex NLS1 1H0707-495 is taken as unequivocal support for the high
spin relativistic reflection picture, as this implies distances of the
source from the disc of $<2R_g$ for a $3\times 10^6M_\odot$ black
hole. The 'simple' NLS1 PG1244+026 shows a much longer lag time of
$\sim 200$~s (Alston, Done \& Vaughan 2014), consistent with the
source being somewhat further from the horizon. 

Lags are derived by comparing variability in a soft and hard band, and
show much more structure than just the reverberation signal. The lags
change as a function of the timescale of variability. The soft leads
the hard for long timescale variability, then this lag decreases with
decreasing variability timescale until it goes
negative (soft lagging) for the fastest fluctuations. This
characteristic soft lead decreasing with frequency was first seen in
the (much higher signal to noise) black hole binaries (BHBs,
e.g. Miyamoto et al 1988; Nowak et al 1999), where it is now
interpreted as the signature of fluctuations propagating through an
inhomogeneous accretion flow (Kotov et al 2001; Arevalo \& Uttley
2006). Slow variability starts at large radii, where the spectrum is
softer. It then propagates down through the entire flow until it gets
to the centre, modulating the hardest spectral region. Faster
variability is only produced closer to the black hole. It has a
shorter distance to travel to reach the inner regions
so takes a shorter time before it
modulates the hardest X-ray component from the centre, i.e. this
characteristic lag pattern is produced by a correlation of
spectrum and variability with radial position. But this also means
that the soft and hard bands both contain some emission from the
softer spectrum produced at large radii, and the harder spectrum
produced at smaller radii. There is just more of the softer (large
radii) spectrum in the soft band than in the hard (and conversely,
more of the harder, small radii spectrum in the hard band than in the
soft). Thus each band contains the same components, but with different
weighting, so most of the variability is perfectly correlated, with
only a very small fraction which is lagged. This is seen in BHBs in a
cross-correlation of hard and soft bands, which always peaks at zero
lag, with just a small asymmetry (Torii et al 2011). The important
point for AGN is that the spectral content of the hard and soft bands
matters, and that the lags are diluted if the same component appears
in both bands (Miller et al 2010; Uttley et al 2014).

Thus lags between light curves in two different energy bands have to be
interpreted with a model of the spectral components which contribute
to each band. Even the 'simple' NLS1 PG1244+026 has a spectrum which
requires multiple components. These can either be modelled by a disc,
low temperature Comptonisation and high temperature Comptonisation,
together with its (weak and only weakly smeared) reflection from a
disc, or with disc, high temperature Comptonisation and its (strong,
and strongly smeared) reflection from a disc (Jin et al 2013), or with
a steep power law (presumably from the jet?) together with high temperature
Comptonisation and its (strong and strongly smeared) reflection from a
disc (Kara et al 2014). Using variability breaks degeneracies, and the
combined constraints from energy spectra, power spectra in hard
and soft bands, high frequency covariance spectra from the 4-10~keV
light curve (the spectrum of the fast variability which correlates with
the 4-10 keV light curve) and the lag-frequency spectrum strongly support
the first model. In this model, slow fluctuations in the disc propagate down
to modulate faster variability in the soft excess (dominated by the low temperature Compton emission), which propagates
down into the high temperature Compton emission from the X-ray corona, which then reflects from and is reprocessed
by the disc. None of the models with strong, strongly smeared
relativistic reflection could adequately match all the spectral-timing
data for this source (Gardner \& Done 2014, hereafter GD14). In particular the fact relativistic reflection contributes strongly to both the hard and soft bands, results in nearly identical hard and soft band power spectra and high coherence between hard and soft bands up to high frequencies, in disagreement with observations of PG1244+026.

This model had no constraints on spin, as the disc component came from
$20-12R_g$, the soft excess from $12-6R_g$ and the coronal size was
assumed to be $6R_g$. The soft X-ray excess is a composite of a
separate low temperature Comptonisation component from the disc (which
carries the required propagation signal) and the mostly thermalised
reprocessed emission from non-reflected X-ray illumination of the
disc. Hence the soft band contains components which carry the soft
lead and the soft lag, but identifying the lag with reprocessing
rather than reflection removes the requirement for extreme
relativistic smearing as the thermalised emission is already smooth,
as seen in the data, quite unlike ionised reflection which has strong
atomic features. NuSTAR data (AKN 120: Matt et al 2014) and long
term X-ray/UV variability (MRK 509: Mehdipour et al 2011; Petrucci et
al 2013; Boissay et al 2014) favour the low temperature Comptonisation
model for the soft X-ray excess in standard broad line Seyfert
1s. This removes the requirement of high spin derived from assuming
that this component is instead from ionised reflection (Patrick et al
2011; see Crummy et al 2006), but we note that the issue is still
controversial, and probably complex (Lohfink et al 2012).

With this model, the lag in PG1244+026 does not require high spin but
neither does it rule out a source at $\sim 6R_g$ above the event
horizon of a high spin black hole, still allowing the range from
'simple' to 'complex' behaviour to be due to the height of the source.
However, the broad band spectral energy distribution {\em does} rule
out a standard disc around a high spin black hole. The observed
optical/UV continuum constrains the mass accretion rate onto the black
hole given its mass, so the total luminosity is only dependent on the
efficiency i.e. on black hole spin. A high spin black hole produces
too much luminosity which is very difficult to hide even in the UV-EUV
interstellar absorption gap (Done et al 2013). Advection is unlikely
to change this conclusion, firstly as this source is not strongly
super-Eddington, and secondly as the most recent global radiative MRI
(but Newtonian) disc simulations of a super-Eddington flow at
$\dot{M}_{in}=20\dot{M}_{Edd}$ show that this is still almost as
radiatively efficient as a standard thin disc due to vertical
convection (Jiang, Stone \& Davis 2014). Thus this source is
most likely low spin, so should never show the extremely deep dips
which are characteristic of the 'complex' NLS1, as a low spin black
hole cannot produce extreme light bending. This is because the steep emissivity index required can only be produced when the illuminating source height is $<2R_g$ (Dovciak et al 2014), which for a low spin BH is below the horizon. This suggests that 'simple'
to 'complex' is more than just the source height, and points instead
to spin as the potential difference between the two classes.  Yet many
NLS1s show transitions between 'simple' and 'complex' behaviour (Gallo
2006; Miller et al 2008) so spin cannot be the origin of the
difference either.

The accretion flow in steady state should only depend on black hole
mass, spin, and mass accretion rate, plus inclination with respect to
the disc which can change its appearance. All the NLS1s have similar
mass and mass accretion rate (within a factor of $\sim 3$) so if the
difference between 'simple' and 'complex' is neither source height nor
spin, then inclination is the last parameter to try. These sources are
all around Eddington, so it seems unlikely that they are actually
described by a flat disc as assumed in the relativistic reverberation
models. The global MRI simulation at $\dot{M}_{in}=20\dot{M}_{Edd}$
showed a complex flow structure, with a large scale height radiation
pressure driven wind from the inner disc carrying away 30\% of the
input mass accretion rate (Jiang et al 2014). This wind is
likely to be less strong in PG1244+026 since this source is only
$\dot{M}_{in}\sim \dot{M}_{Edd}$.  UV line driving is unlikely to
help since the low mass and high mass accretion rate
of NLS1s mean that their disc is too hot for its photosphere to have
the required opacity to UV line transitions (Hagino et al 2014).

Hence strong mass loss is not expected in the NLS1s, but some
turbulent, clumpy, failed Eddington wind in the inner disc could
easily form (Jiang et al 2014). High inclination angles have a high
probability of a clump intersecting the line of sight, while low
inclination angles are mostly free of obscuration, giving a potential
mapping from complex to simple NLS1 as a function of inclination.
The spectral signature of this time dependent absorption could be
complex, depending on the amount of source occulted and the ionisation
state of the absorber.

Here we explore the effect of occultations on the lag-frequency
behaviour. We show that occultations produce a soft lead when
relativistic Doppler effects are taken into account. We couple
orbiting clump occultations with the full spectral timing model
developed in GD14 to explain the properties of the simple NLS1
PG1244+026. We show that multiple occultations of this intrinsic
propagation and reprocessing/reflection model can replicate the change
in observed lag-frequency properties from a simple to a complex NLS1. In
particular the soft lead resulting from occultations dilutes the
maximum measured reverberation lag and shifts it to higher
frequencies, as observed. 

Many issues still remain for this to be a full model of the complex
NLS1. Nevertheless, this shows a promising alternative 
geometry for the complex NLS1.

\begin{figure*} 
\centering
\begin{tabular}{l|r}
\leavevmode  
\includegraphics[width=8cm]{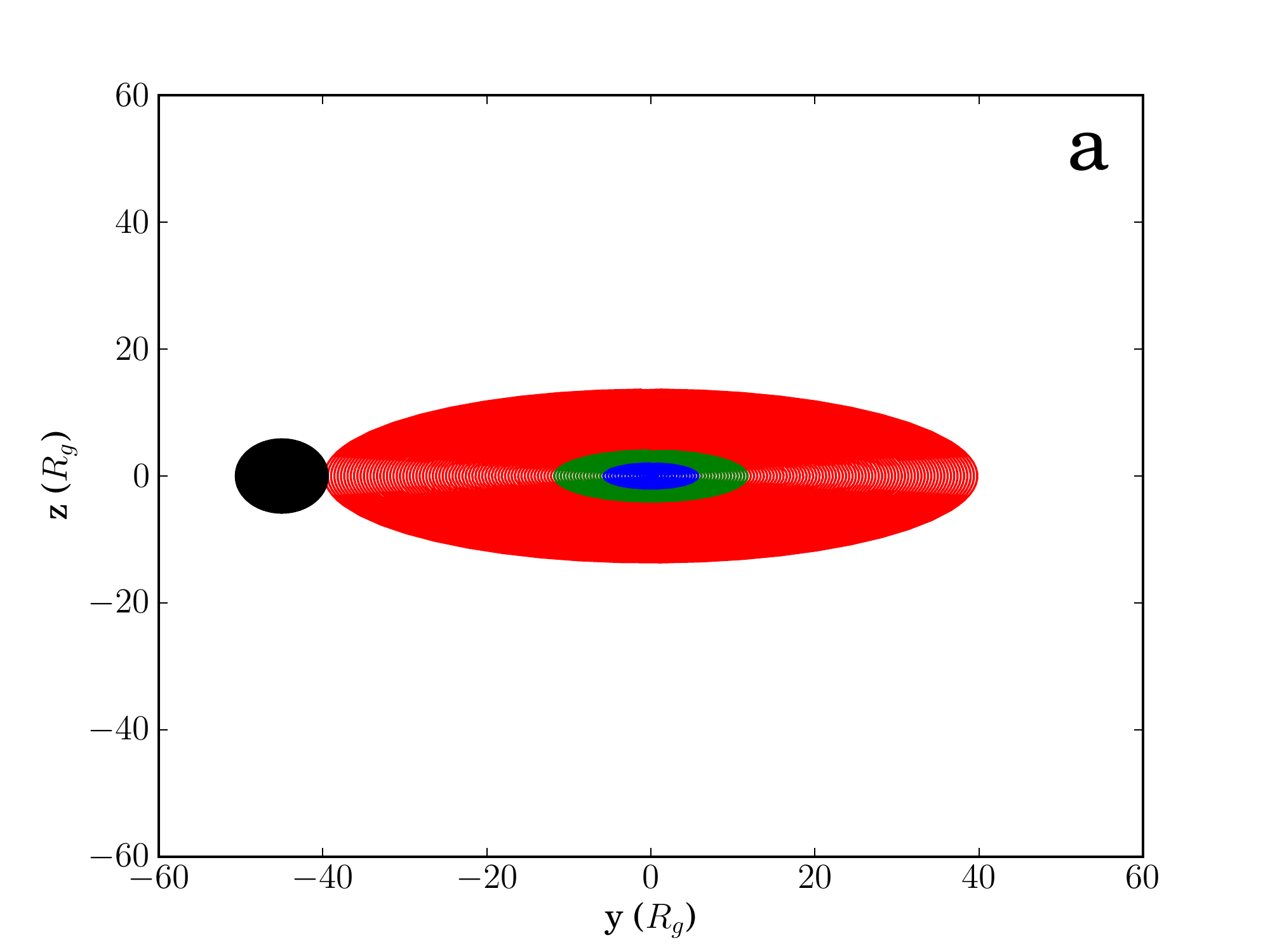} &
\includegraphics[width=8cm]{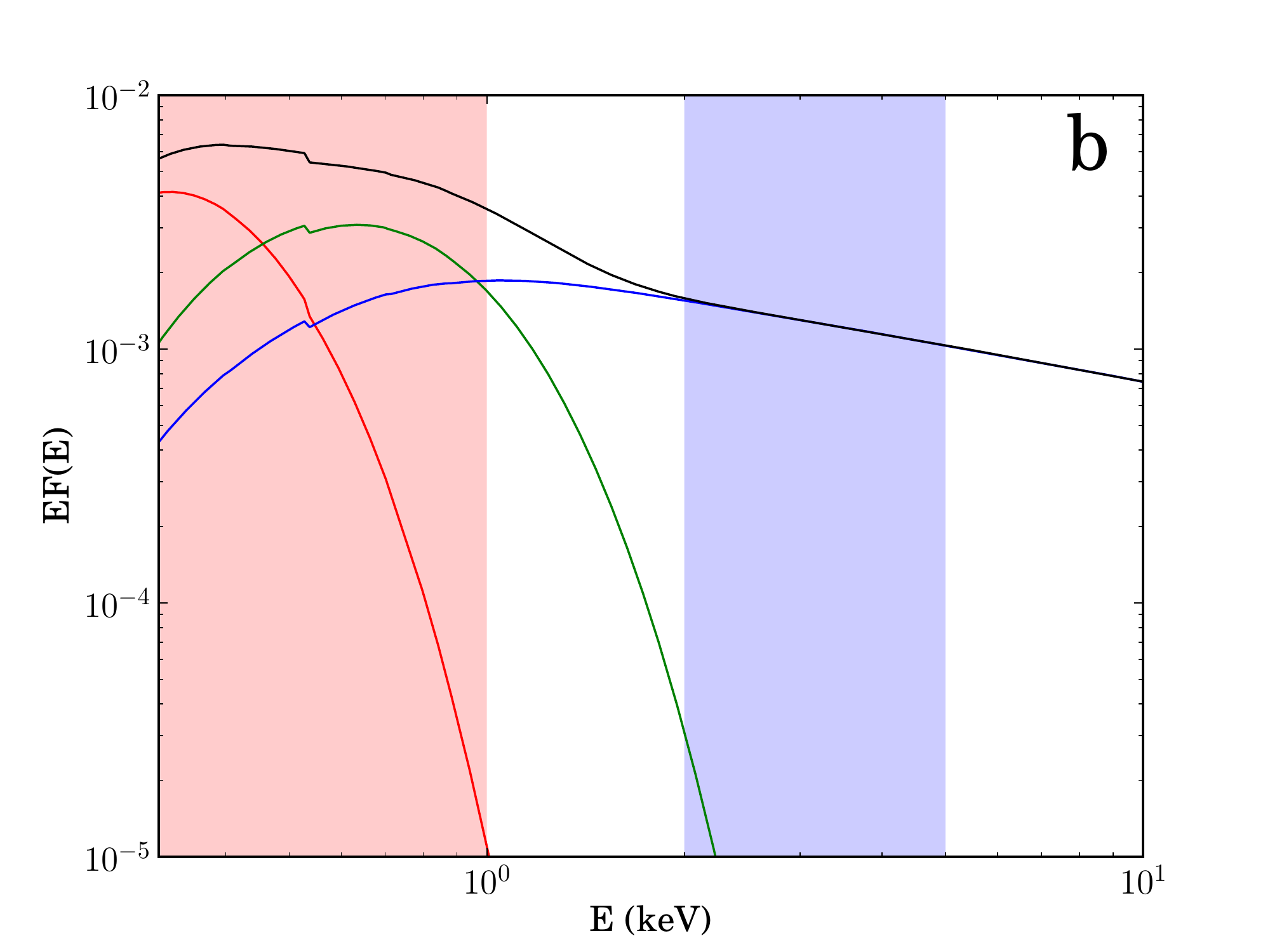} \\
\end{tabular}
\caption{a). Physical location of accretion flow model components, as viewed at high inclination ($i=70^\circ$): disc (red), soft excess (green), corona (blue), obscuring cloud (black). For simplicity we have approximated the geometry of the accretion flow as a flattened disc. b). Spectrum emitted by each component, with total spectrum shown in black, and hard and soft bands shaded in blue and red respectively.}
\label{fig1}
\end{figure*}

\section{Linear Occulation}

We choose to model the underlying accretion flow using the separate
soft excess model found by GD14 to describe the spectral timing
properties of the simple NLS1 PG1244+026. We assume the accretion flow
consists of three components. The outermost radii form a standard
accretion disc. This is truncated at some radius ($R_{cor}\sim
20R_g$). The remaining gravitational energy liberated between
$R_{cor}$ and the innermost stable circular orbit ($R_{isco}$) is used
to power the remaining two components: the soft excess and the
corona. Below $R_{cor}$ material is unable to thermalise completely
and form a cool accretion disc, perhaps due to a larger scale height
from the photosphere lifting to form the (failed) wind. Instead some
of the electrons emit via optically thick Comptonisation of the
cooler disc seed photons. This optically thick Compton emission adds
to the spectrum at low energies, producing an excess of soft X-rays,
hence we call the physical region producing this emission the 'soft
excess'. An optically thin corona extends above the soft excess at the
very smallest radii. This corona contains hot electrons at lower
density and higher energy than the soft excess and provides a source
of optically thin Comptonisation, using seed photons from the cooler
soft excess region (Jin et al 2013). 

Fig. 1a shows the physical locations of each model component and
Fig. 1b shows their contributions to the total X-ray spectrum. For
simplicity we approximate the geometry of the accretion flow as a
flattened disc. We model the wind as a series of individual clouds
which transit the flow (from left to right in Fig. 1a, co-rotating
with the flow) and obscure the intrinsic emission. We assume identical
spherical clouds. The model has five free parameters: cloud transit
time, cloud radius, cloud number density, transit latitude and
cloud ionisation. In the following sections (\S2-3) we
investigate the effect of occultations with no intrinsic variations in
luminosity of the accretion flow components, i.e. we assume constant
flux from the underlying accretion flow. In all cases we assume an
inclination angle of the accretion flow with respect to the line of
sight of $70^\circ$ and black hole mass of $10^7M_{\odot}$ and define
a soft band from $0.3-1$keV and hard band from $2-5$keV (shaded in red
and blue respectively in Fig. 1b).

\begin{figure*} 
\centering
\begin{tabular}{l|r}
\leavevmode  
\includegraphics[width=8cm]{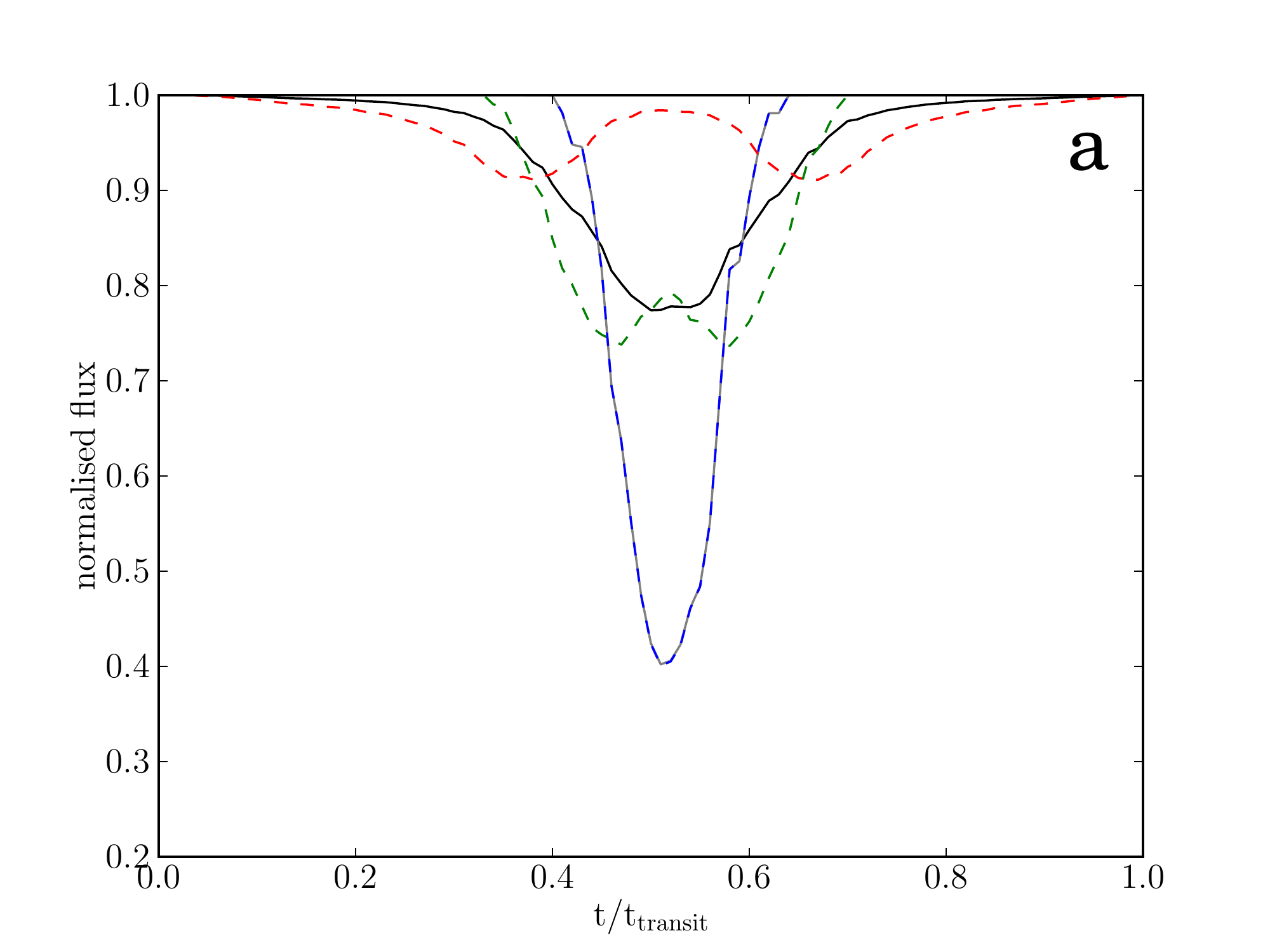} &
\includegraphics[width=8cm]{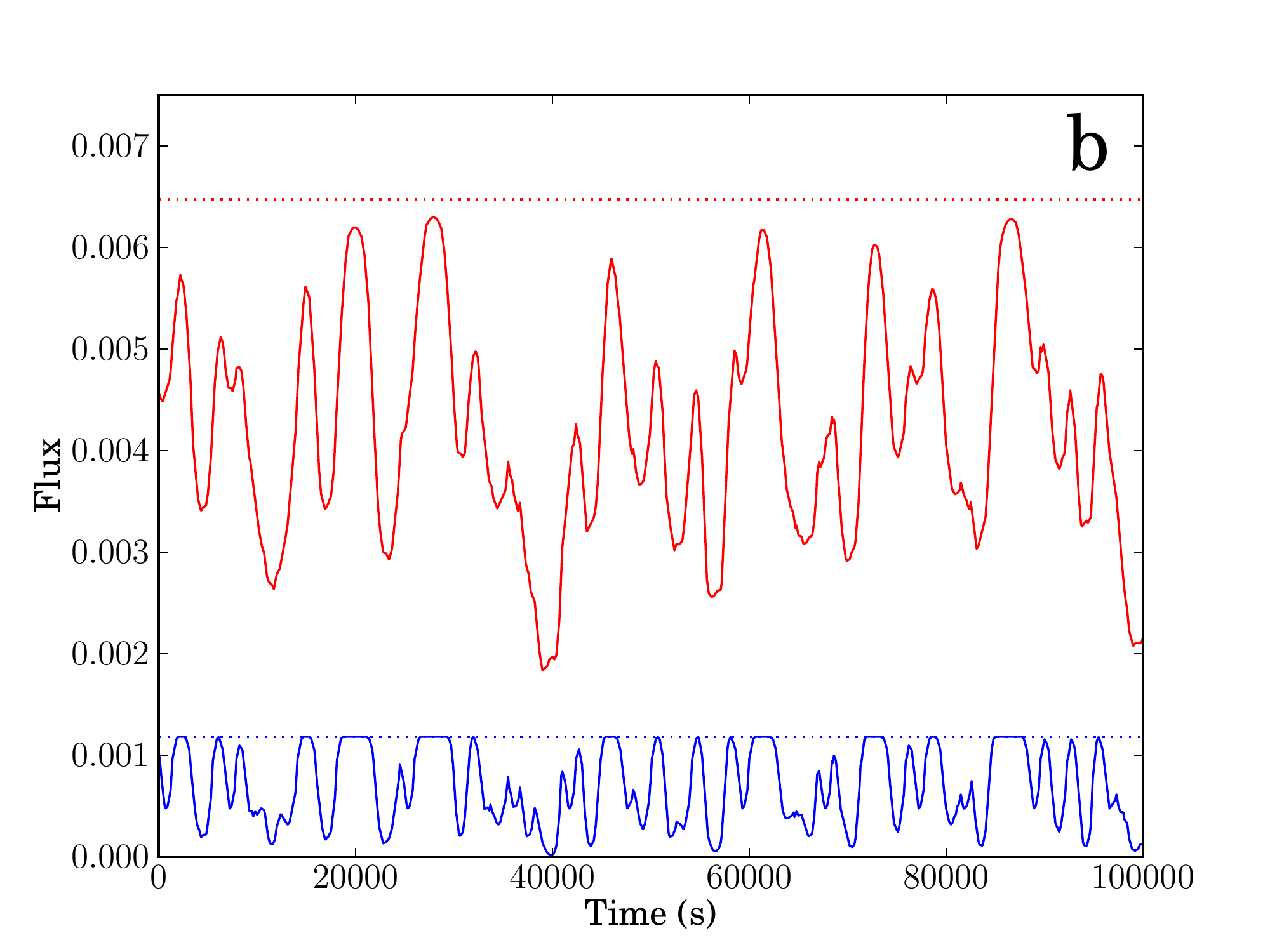} \\
\includegraphics[width=8cm]{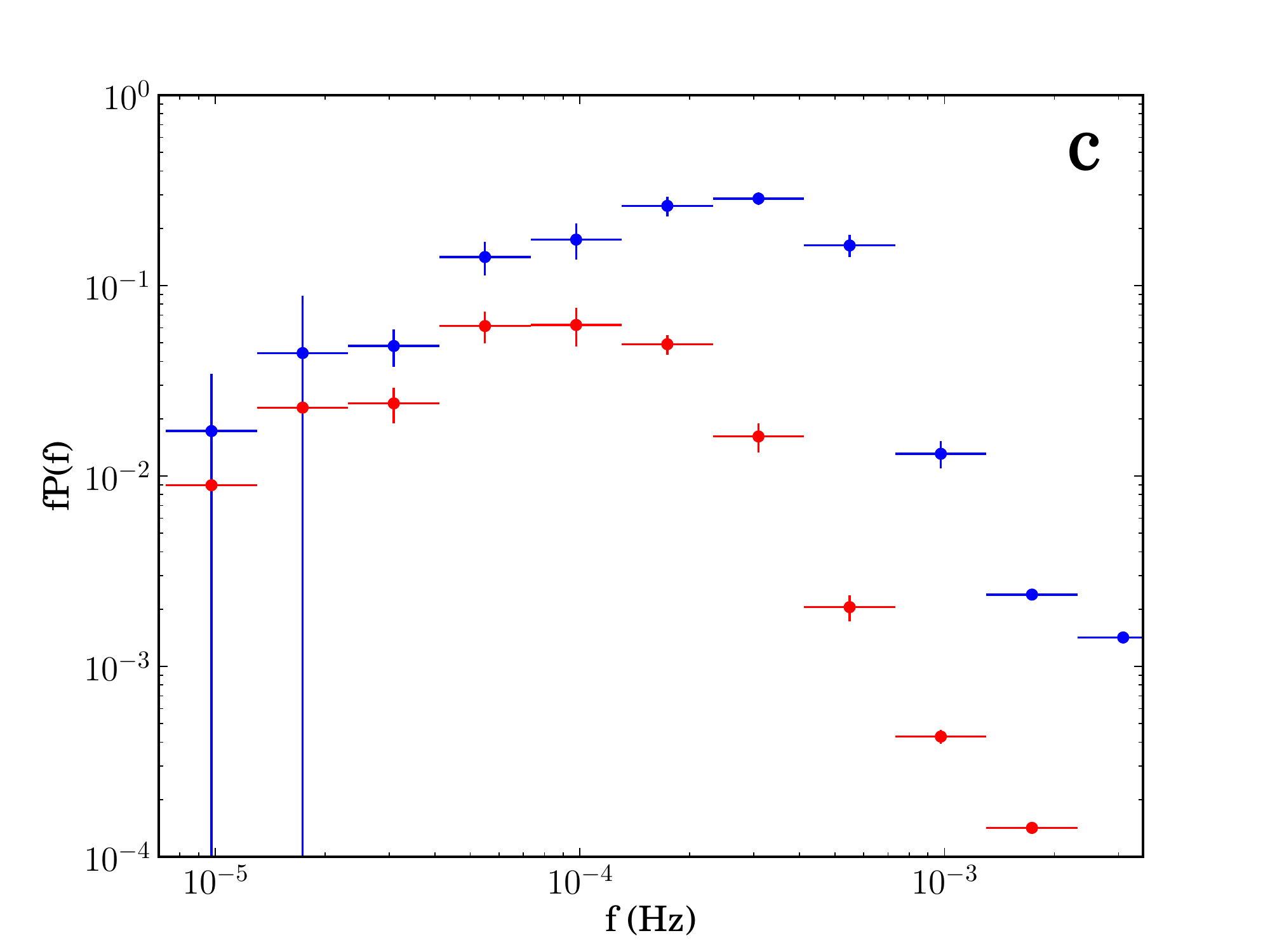} &
\includegraphics[width=8cm]{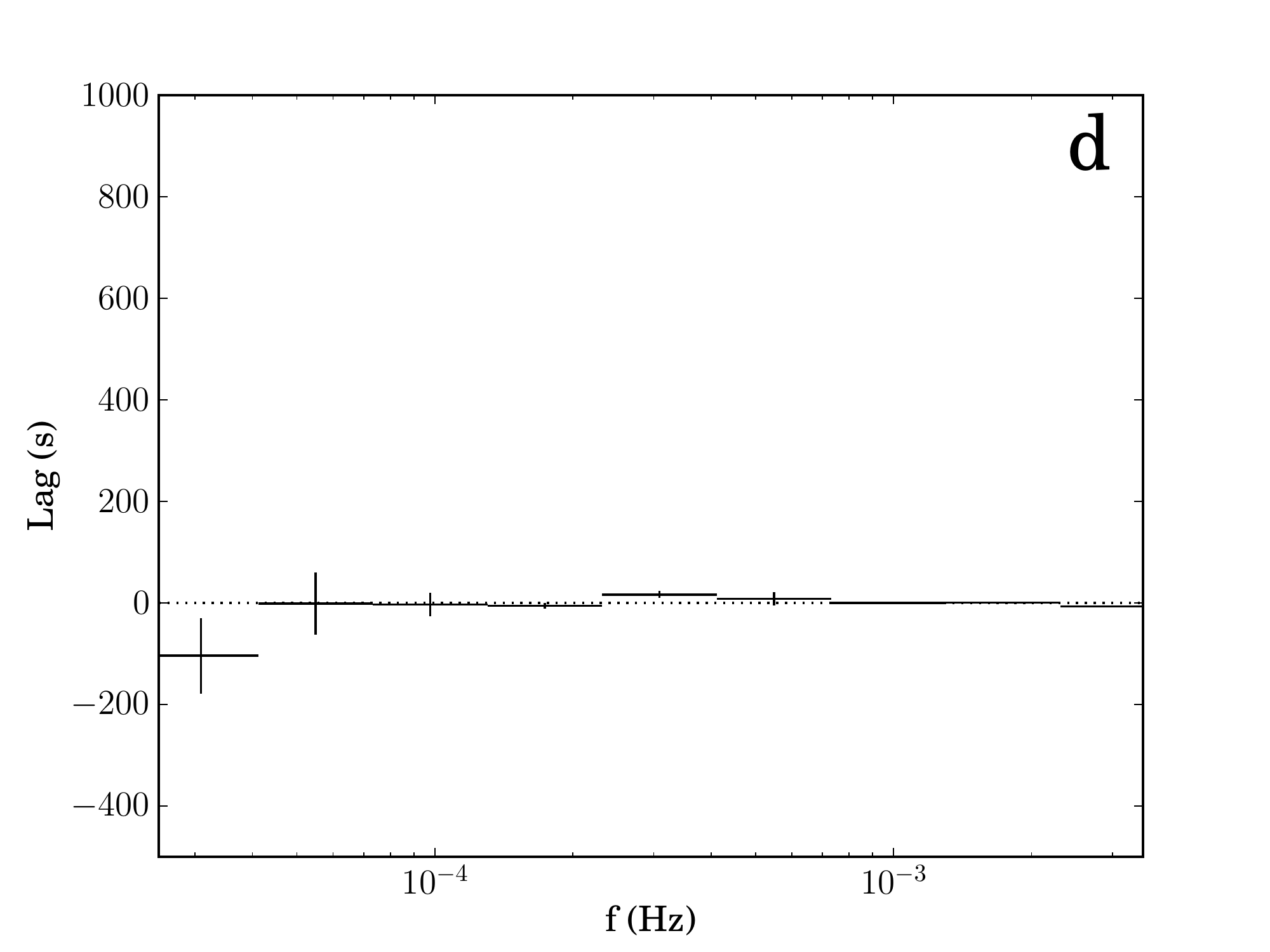} \\
\end{tabular}
\caption{Model occulting a static accretion flow. a). Fractional flux drop for a single occultation as a function of total transit time: disc (dashed red), soft excess (dashed green), corona (dashed blue), hard band ($2-5$keV, solid grey; same shape as coronal flux drop), soft band ($0.3-1$keV, solid black). b). Sample hard (blue) and soft (red) model light curves, showing effect of occultations. Dotted lines show unobscured flux level. c). Hard and soft band power spectra (blue and red respectively). d). Lag-frequency spectrum between hard and soft bands.}
\label{fig2}
\end{figure*}

\begin{figure*} 
\centering
\begin{tabular}{l|r}
\leavevmode  
\includegraphics[width=8cm]{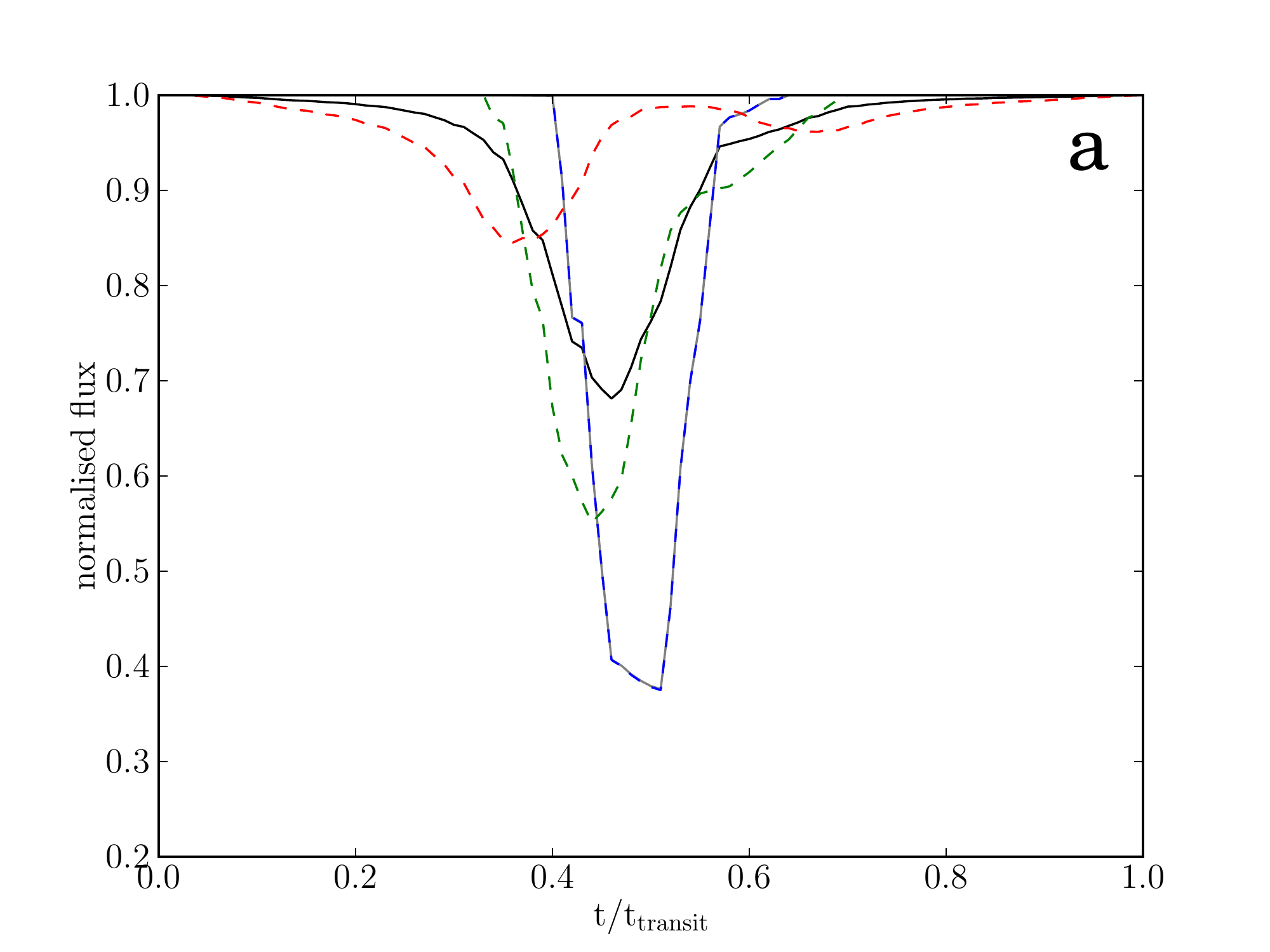} &
\includegraphics[width=8cm]{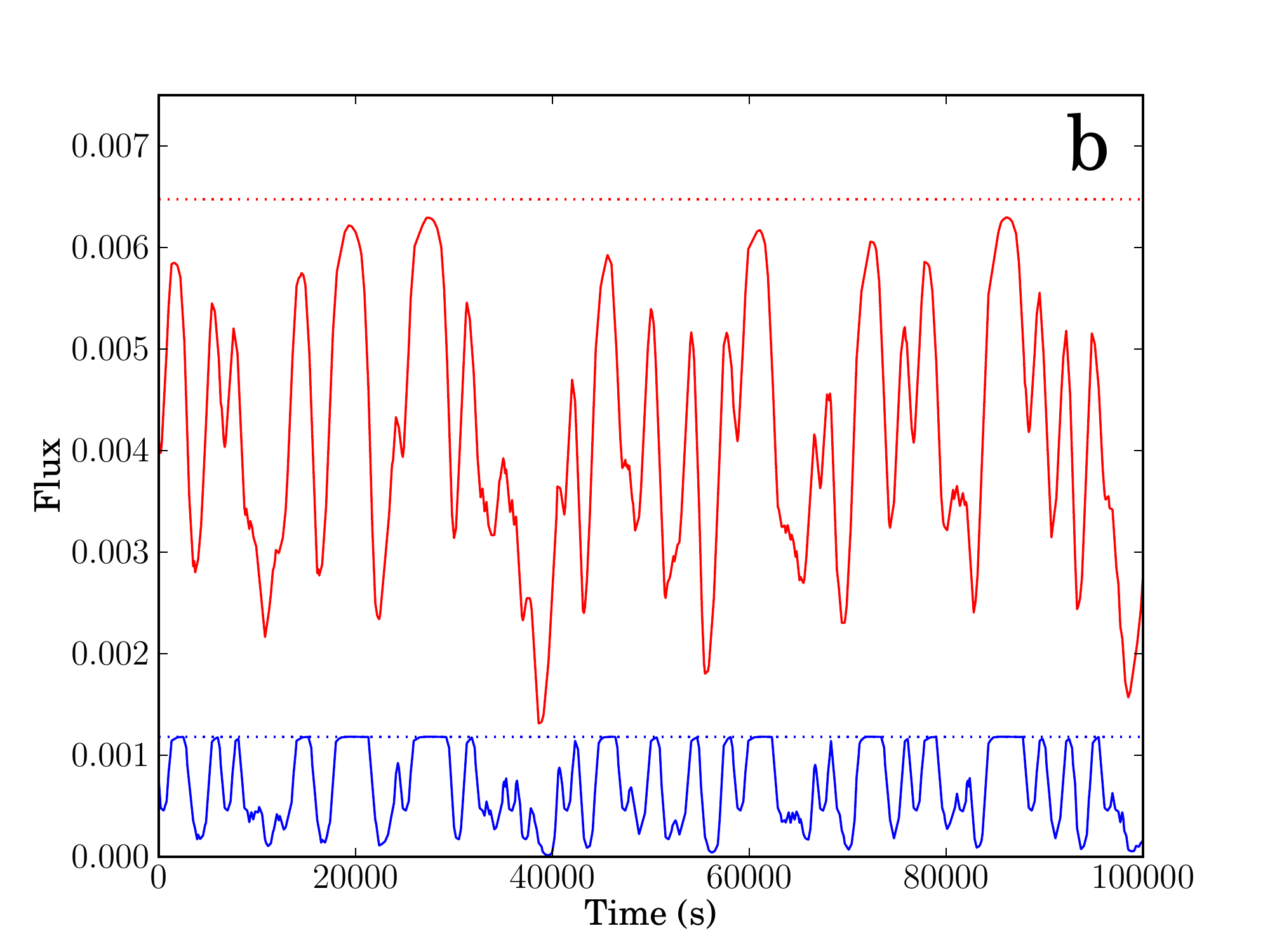} \\
\includegraphics[width=8cm]{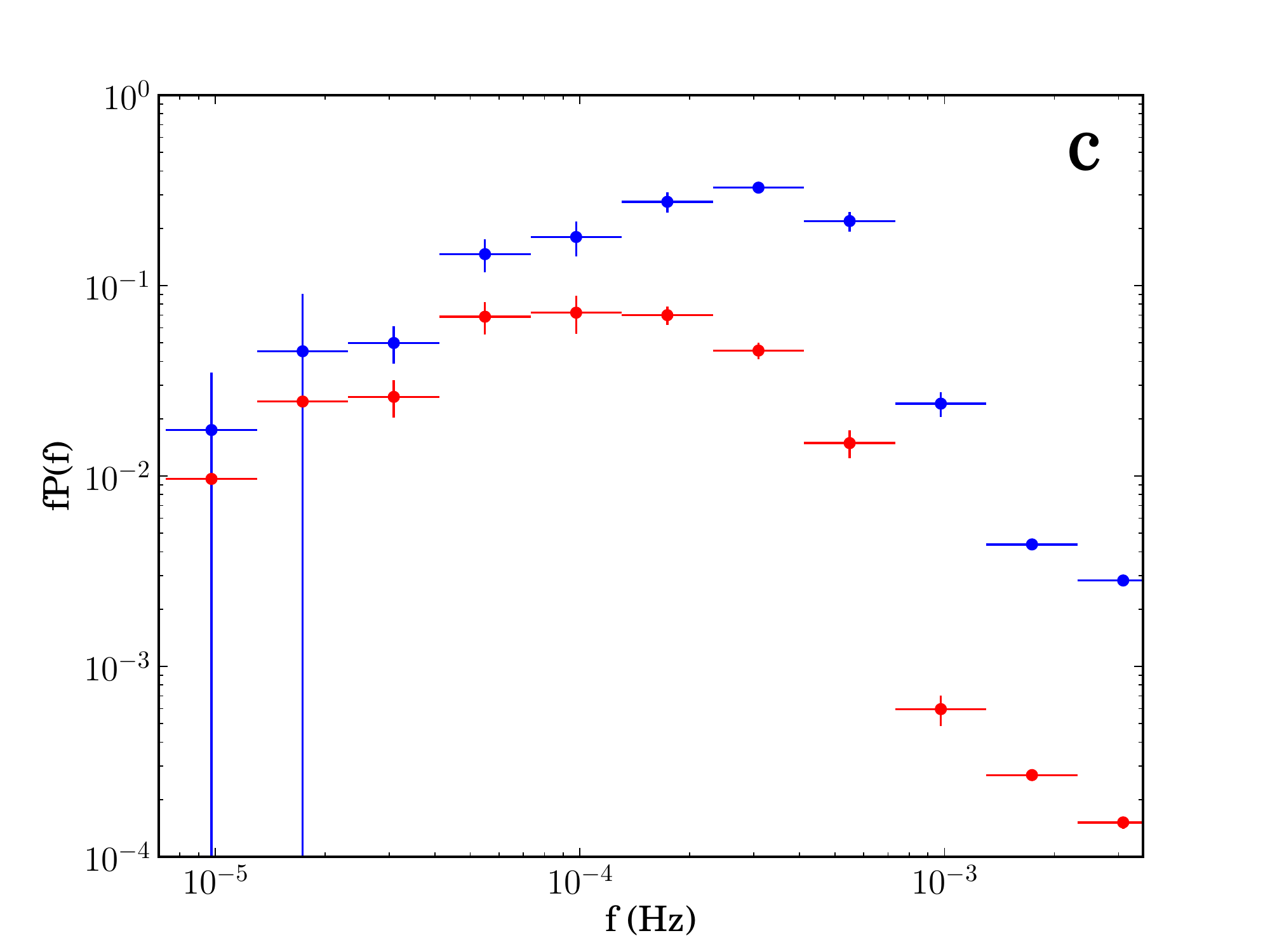} &
\includegraphics[width=8cm]{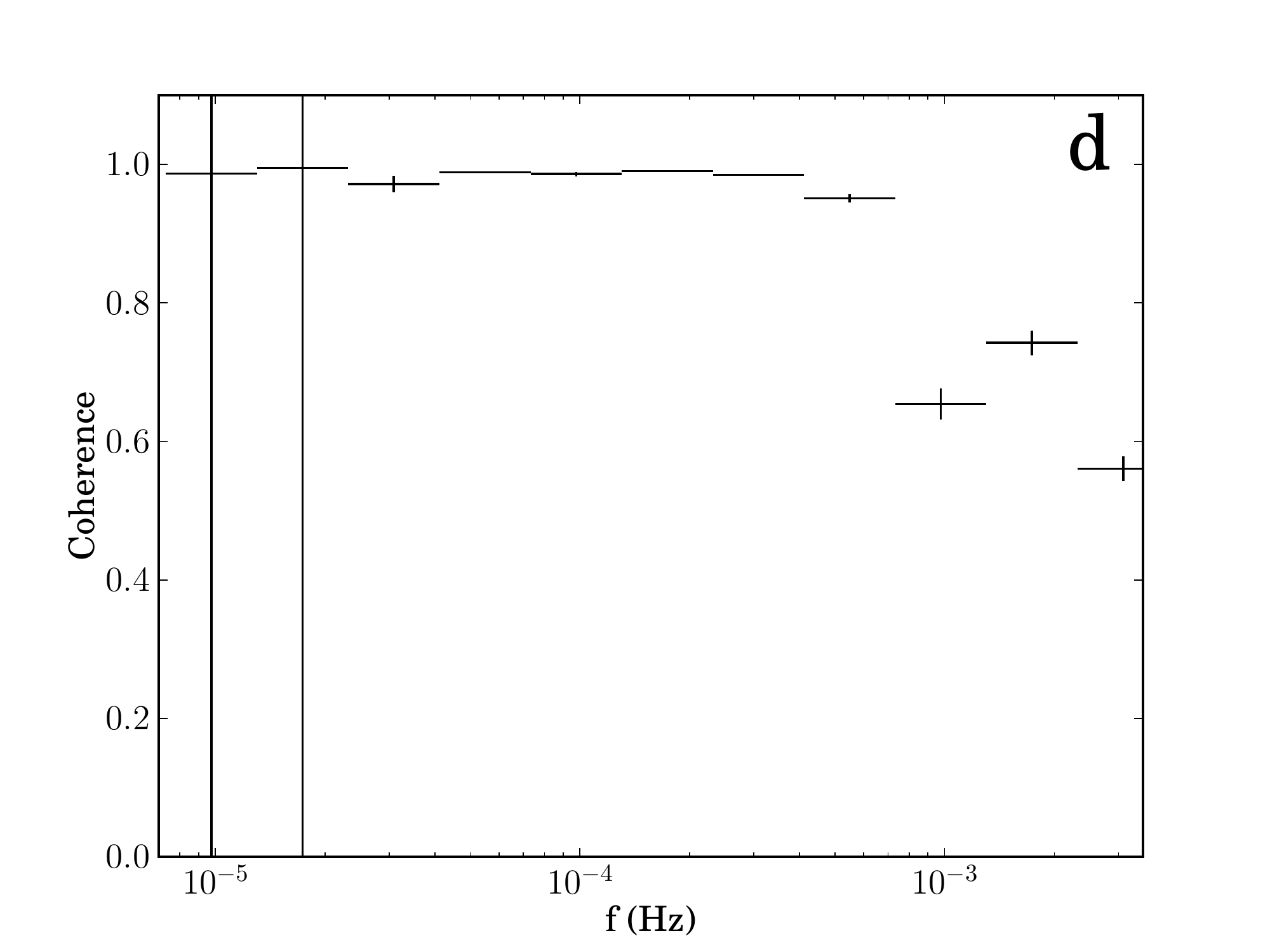} \\
\includegraphics[width=8cm]{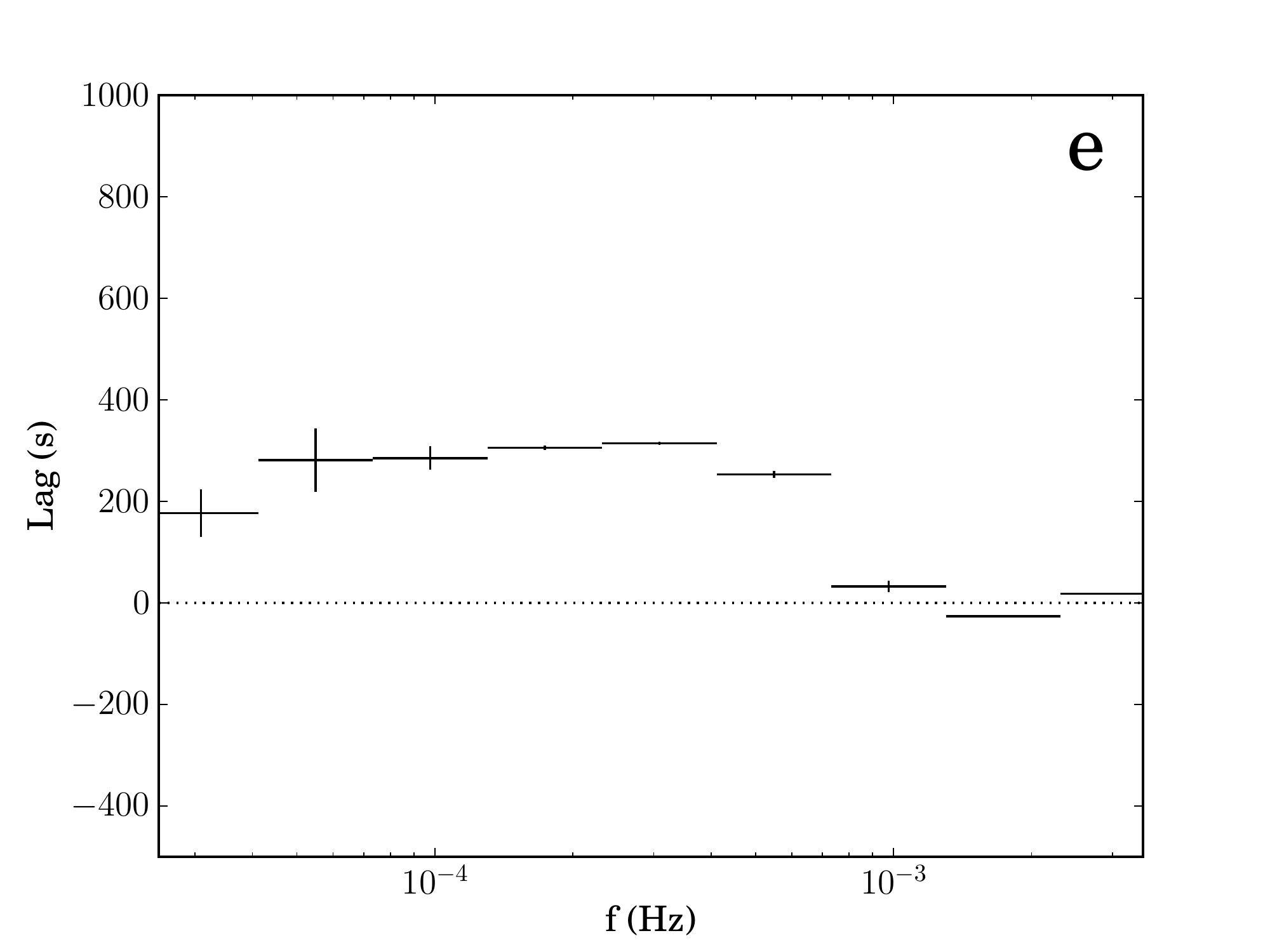} &
\includegraphics[width=8cm]{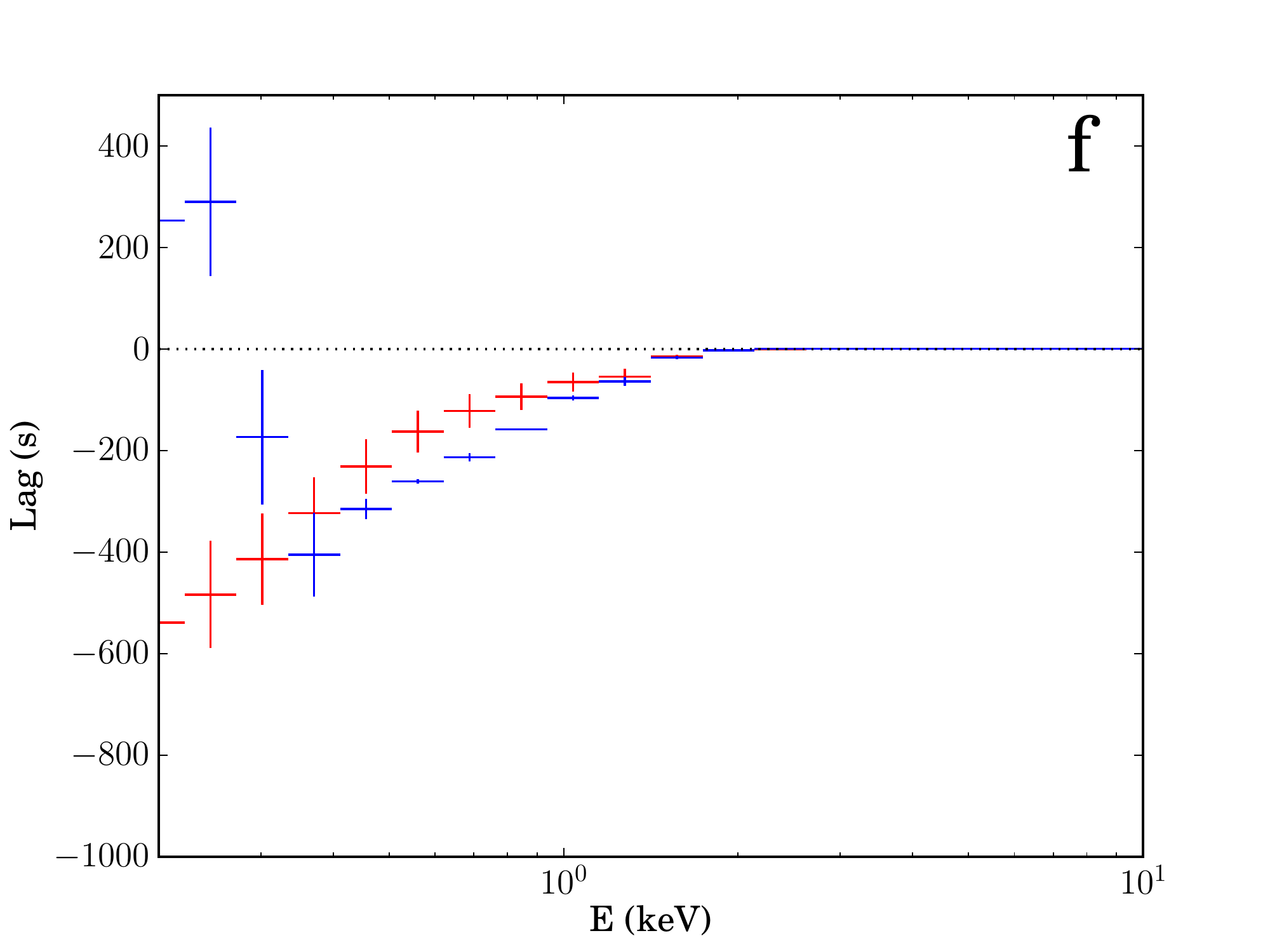} \\
\end{tabular}
\caption{Occultation model including effects of Doppler boosting. a). Fractional flux drop for a single occultation as a function of total transit time: disc (dashed red), soft excess (dashed green), corona (dashed blue), hard band ($2-5$keV, solid grey; same shape as coronal flux drop), soft band ($0.3-1$keV, solid black). b). Sample hard (blue) and soft (red) model light curves, showing effect of occultations. Dotted lines show unobscured flux level. c). Hard and soft band power spectra (blue and red respectively). d). Coherence between hard and soft energy bands. e). Lag-frequency spectrum between hard and soft bands. f). Lag-energy spectrum calculated using $2-5$keV reference band. Red points show energy spectrum of lag at low frequencies ($2.3\times10^{-5}-7.3\times10^{-5}$Hz), blue points show the lag-energy spectrum at high frequencies ($2.3\times10^{-4}-7.3\times10^{-4}$Hz).}
\label{fig3}
\end{figure*}

\subsection{Effect of Doppler Boosting}

Fig. 2a shows the fractional flux drop as a single cloud passes across a static accretion flow. We assume the cloud reduces the observed flux by $e^{-\tau}$, where the optical depth $\tau=1$ and does not vary with the energy of the incident radiation, i.e. the material is completely ionised. 

The dashed red line shows the drop in disc flux. Since the apparent size of the disc is much larger than the cloud the drop in flux is small ($<10\%$). We assume a radial emissivity profile for the disc of $\varepsilon(r)\propto r^{-3}$, so when the cloud occults the outer parts of the disc the flux drop is small. The flux drop increases as the transit progresses and the cloud begins to occult the inner brighter disc radii. The disc flux recovers during the middle of the transit, as the cloud passes over the innermost regions occupied by the corona and soft excess, and then drops again as the cloud crosses the far side of the disc. The dashed green line shows the drop in flux from the soft excess. This shows similar behaviour, but more centrally concentrated, since the soft excess region is smaller. The corona shows the biggest drop in flux (dashed blue line), being a similar size to the occulting cloud. The solid grey and black lines show the flux drop in the hard and soft bands respectively. The hard band is dominated by emission from the corona and hence shows a greater flux drop than the soft band, which is dominated by emission from the disc and soft excess. 

Fig. 2b shows sample hard and soft band light curves (in blue and red
respectively), showing the effect of multiple occultations. Dotted lines show
the unobscured flux levels. The light curves are generated by
allocating each cloud a random start time for its transit of the
disc. Each cloud has a radius of $5R_g$ and takes $T_{tr}=10^4$s to
cross the accretion flow (from left to right as seen in Fig. 1a). On
careful inspection of the light curves it can be seen that the width
of the occultations is narrower in the hard band than the soft band,
due to the smaller physical size of the corona, which is the main
contributor of hard band flux.

Fig. 2c shows the corresponding power spectra. The occultations add
power to the light curve at a frequency that is related to the transit
time. In this simulation all clouds were given a transit time of
$T_{tr} = 10^4$s, corresponding to a frequency of $10^{-4}$Hz. The
power in the hard band peaks at a slightly higher frequency and
greater amplitude. This is a direct result of the the shape of the
flux drops shown in Fig. 2a. The more compact coronal emission
experiences a narrower, deeper flux drop than the more extended soft
band components, hence occultations add more power to the hard band
and at higher frequencies. The width of the hard band flux drop is
$\sim0.2T_{tr}\sim2\times10^3$s, which corresponds to a frequency of
$5\times10^{-4}$Hz. Hence the power drops off sharply above
$5\times10^{-4}$Hz. Nevertheless there is a low power tail extending
to higher frequencies in both bands. A single occultation cannot add
power at these frequencies. This power comes from the superposition of
occultations. The hard and soft band light curves in Fig. 2b show that
multiple occultations close together can add variability on
timescales much shorter than that of an individual transit.

Fig. 2d shows the lag as a function of frequency between the hard and soft bands (calculated following Nowak et al. 1999; see also appendix GD14). There is no lag at any frequency. This is because, even though the soft band flux drop is wider than the hard band, they are both symmetric around a common centre. Even though the soft band flux drops before the hard, the hard flux then recovers before the soft with the same time delay, cancelling out any net lag. This is the case for a stationary disc. 

However the accretion flow is not stationary. Material should be rotating at the Keplerian frequency. As a consequence material on one side of the flow is travelling towards the observer and Doppler boosted, while on the other side the emission is deboosted. Fig. 3 shows the resulting flux drops, light curves, power spectra and lags now including the effect of this Doppler boosting. 

Fig. 3a shows that the flux drop in each component is now no longer symmetric. Doppler boosting means that the approaching side of the accretion flow appears brighter than the receding side. We assume the occulting clouds are co-rotating with the flow. The approaching side of the flow, which now contributes a greater fraction of the total flux, is occulted first. Hence the first half of the transit shows a much stronger flux drop. The receding side of the flow contributes much less flux to the total spectrum, hence there is a much smaller flux drop during the second half of the transit. This effect is more noticeable in the more extended components - the disc and soft excess - and most noticeable in the soft excess, where the smaller radii give faster radial velocities and stronger Doppler boosting/deboosting. The flux drops in the hard and soft bands (grey and black solid lines) are consequently skewed towards the first half of the transit, with the soft band being more strongly skewed. On closer inspection of Fig. 3a, it can be seen that the soft band flux now drops before the hard flux {\emph{and recovers}} before the hard flux. The occultations themselves have introduced a lag between the hard and soft energy bands. 

Fig. 3e shows the lag as a function of frequency between the hard and soft energy bands. The only source of variability in the light curves are the occultations, we keep the intrinsic flux from the accretion flow constant. By including the effects of Doppler boosting, the occultations have introduced a soft lead of $\sim300$s (a positive lag value indicates the soft band leading the hard). This lead remains roughly constant for the frequency range over which the occultations introduce power into the light curves ($\sim5\times10^{-5}-5\times10^{-4}$Hz). 

Fig. 3f shows the lag as a function of energy. This is constructed by choosing a reference band (in this case the hard band, $2-5$keV) and then dividing the spectrum into a series of energy bins. The flux in each energy bin is summed up as a function of time to create a light curve for that energy bin. The light curve of the energy bin is then compared with the reference band light curve and the value of the lag between the two is computed as a function of frequency. In Fig. 3f we plot the value of the lag from each energy bin for two frequency ranges: low frequency (red points, $2.3\times10^{-5}-7.3\times10^{-5}$Hz) and high frequency (blue points, $2.3\times10^{-4}-7.3\times10^{-4}$Hz). This gives the energy spectrum of the lag at that frequency. For each energy bin, a negative lag value implies that energy bin leads the hard reference band. 

At low frequencies, energy bins below $1$keV lead the hard reference band, with a lead that increases as the energy of the bin decreases (red points, Fig. 3f). These are the energies at which the disc and soft excess dominate. The soft excess emission peaks at $\sim0.5$keV, giving way to the disc at lower energies. Since the clouds occult the outermost components first, crossing first the disc and then the soft excess before passing in front of the corona, the low energy disc emission shows the strongest lead, giving way to a slightly shorter lead from the soft excess at smaller radii and higher energies. Above $1$keV the emission is dominated by the corona. These are the energies also covered by the reference band ($2-5$keV). Hence the lag of the energy bins with respect to the reference band tends to zero at high energies ($>1$keV). 

This pattern of soft leads (at low frequencies) is a signature normally associated with propagation; low frequency fluctuations are generated in the cooler outer components and propagate down to the hotter smaller radii which produce the high energy emission. Yet in this scenario we have produced soft leads simply by the movement of absorbing clouds, i.e. occultations affect the lag-frequency and lag-energy spectra the same way as propagation, by introducing soft leads. 

Soft leads due to propagation of fluctuations are generally confined to low frequencies, since large radii only generate slow fluctuations. The blue points in Fig. 3f show that occultations can continue producing strong soft leads up to much higher frequencies. This is because the strength and frequency of the soft leads are no longer determined by the properties of the accretion flow but by the properties of the transiting clouds. However, comparison of the red and blue points in Fig. 3f shows that, while all energy bins below 1keV show a soft lead at low frequencies, at high frequencies the two lowest energy bins ($<0.3keV$) switch from a soft lead to a soft lag. These two energy bins are dominated by disc emission. High frequency variability results from short timescale features in the flux drops shown in Fig. 3a, implying this soft lag confined to very low energy comes from the cloud covering the deboosted side of the disc after covering the corona.

For completeness we also show the coherence between hard and soft energy bands (Fig. 3d), where $1$ is perfect coherence between the two bands and $0$ is incoherence. The coherence remains high up to high frequencies, since occultation is the only source of variability in the two light curves and is common to both. The coherence drops off above $\sim 5\times10^{-4}$Hz, where the variability power introduced by the occultations also drops off. $5\times10^{-4}$Hz corresponds roughly to the width of the hard band flux drops ($\sim 0.2T_{tr} \sim 2\times10^3$s).

\begin{figure*} 
\centering
\begin{tabular}{l|r}
\leavevmode  
\includegraphics[width=8cm]{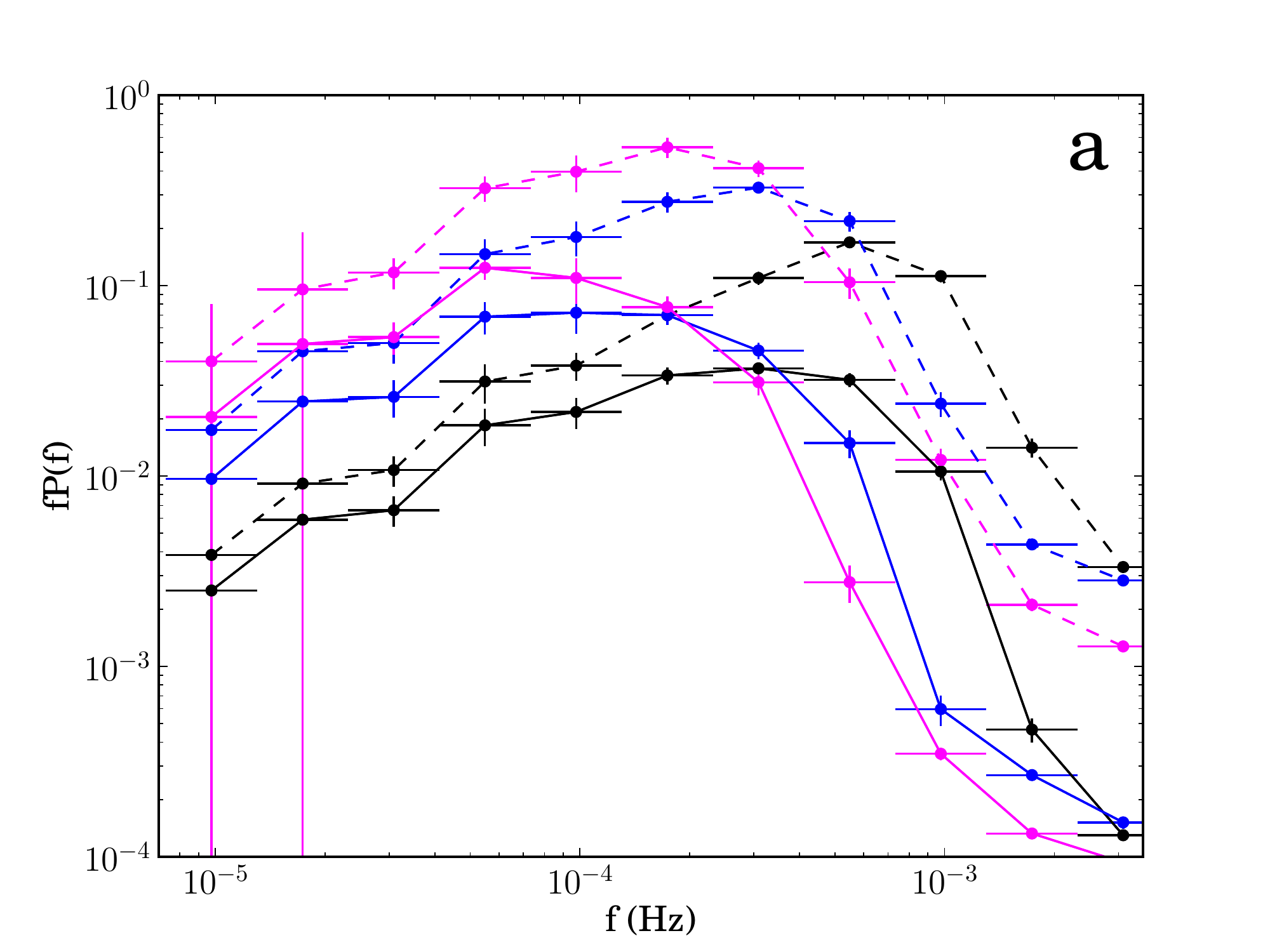} &
\includegraphics[width=8cm]{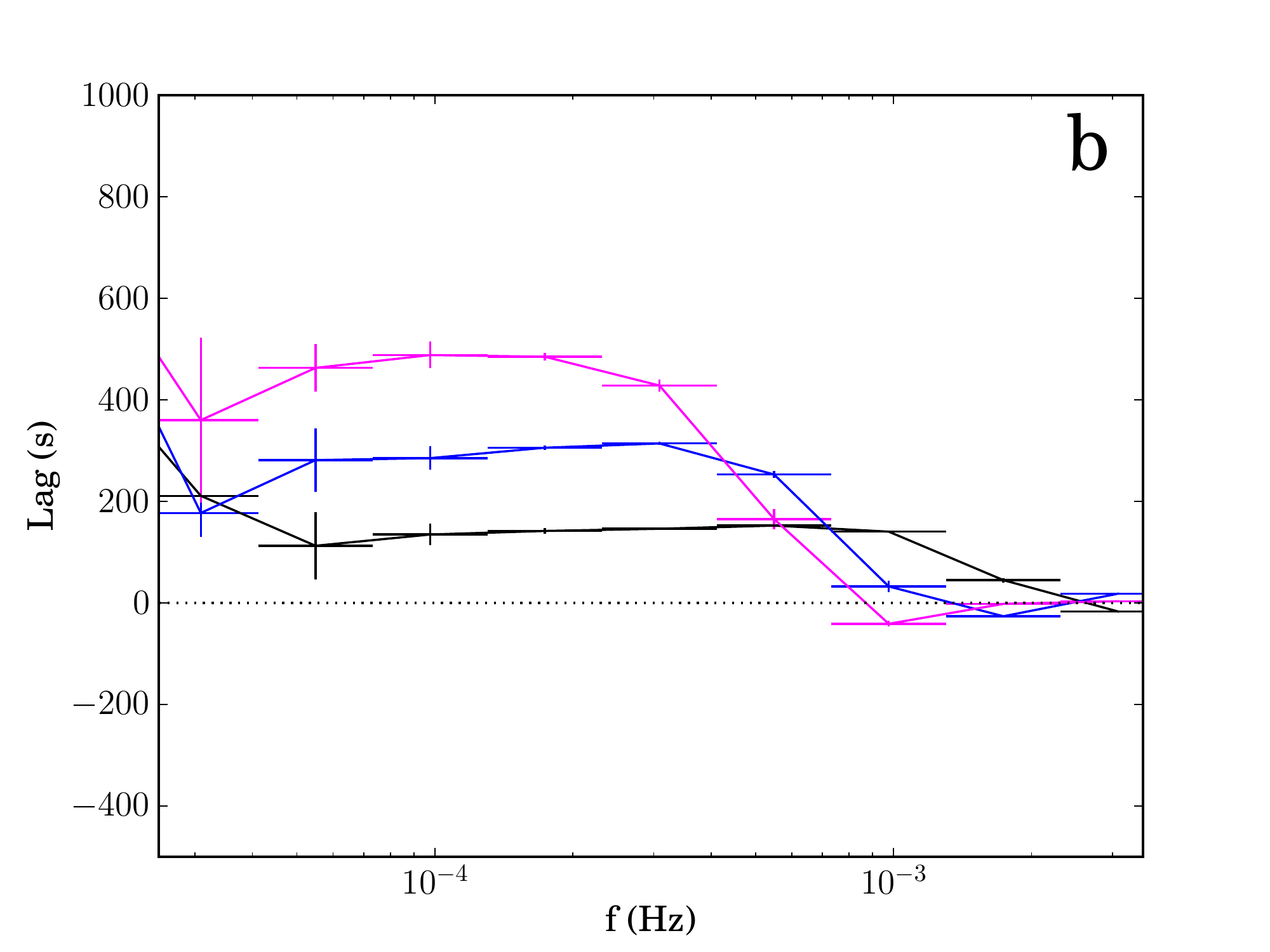} \\
\end{tabular}
\caption{Effect of increasing cloud transit time on a). power spectrum and b). lag-frequency spectrum, for $T_{tr}=5\times10^3$ (black), $10^4$ (blue) and $1.5\times10^4$s (magenta). Solid lines show soft band ($0.3-1$keV) power spectra, dashed lines show hard band ($2-5$keV) power spectra.}
\label{fig4}

\end{figure*}
\begin{figure*} 
\centering
\begin{tabular}{l|r}
\leavevmode  
\includegraphics[width=8cm]{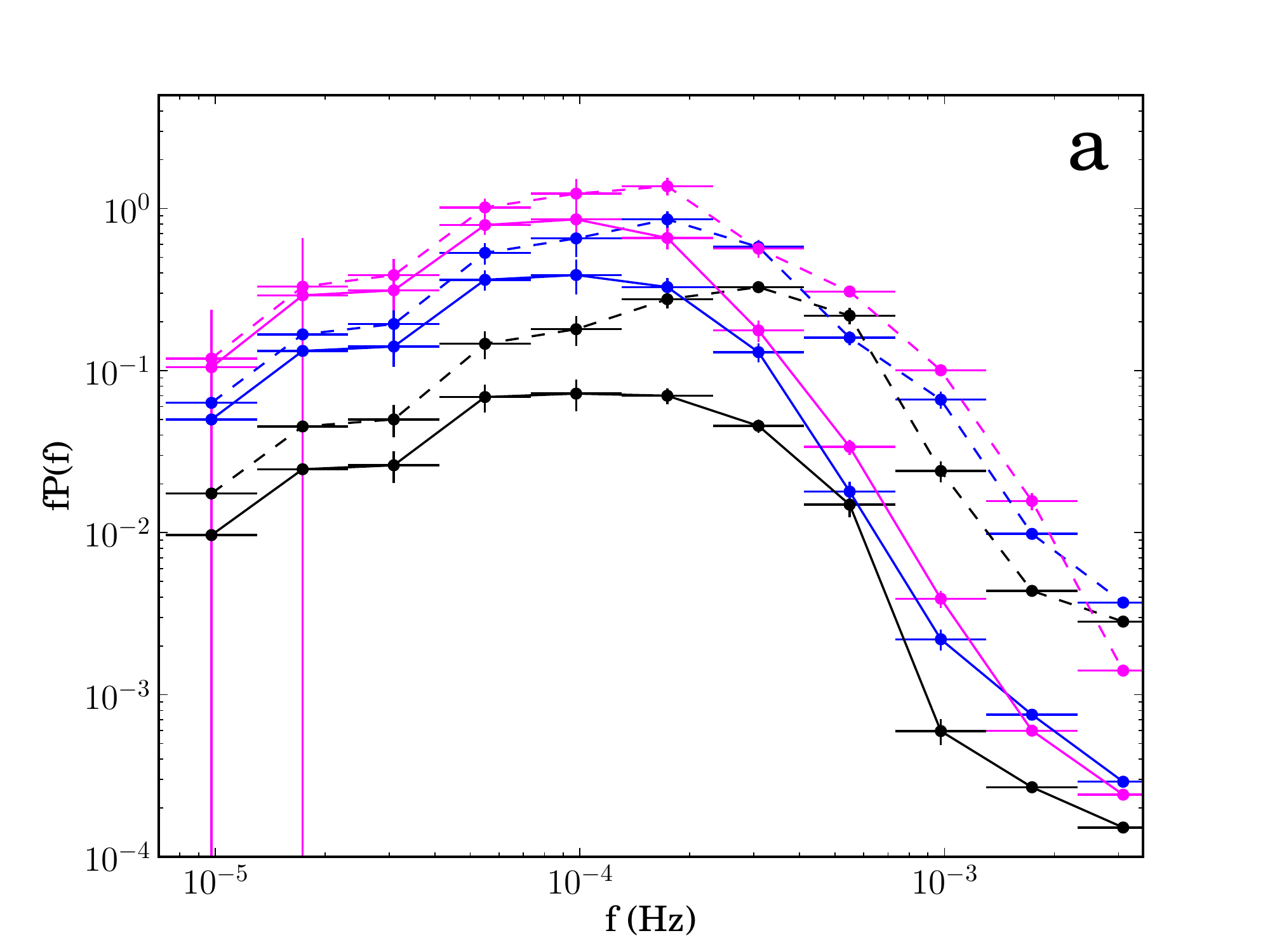} &
\includegraphics[width=8cm]{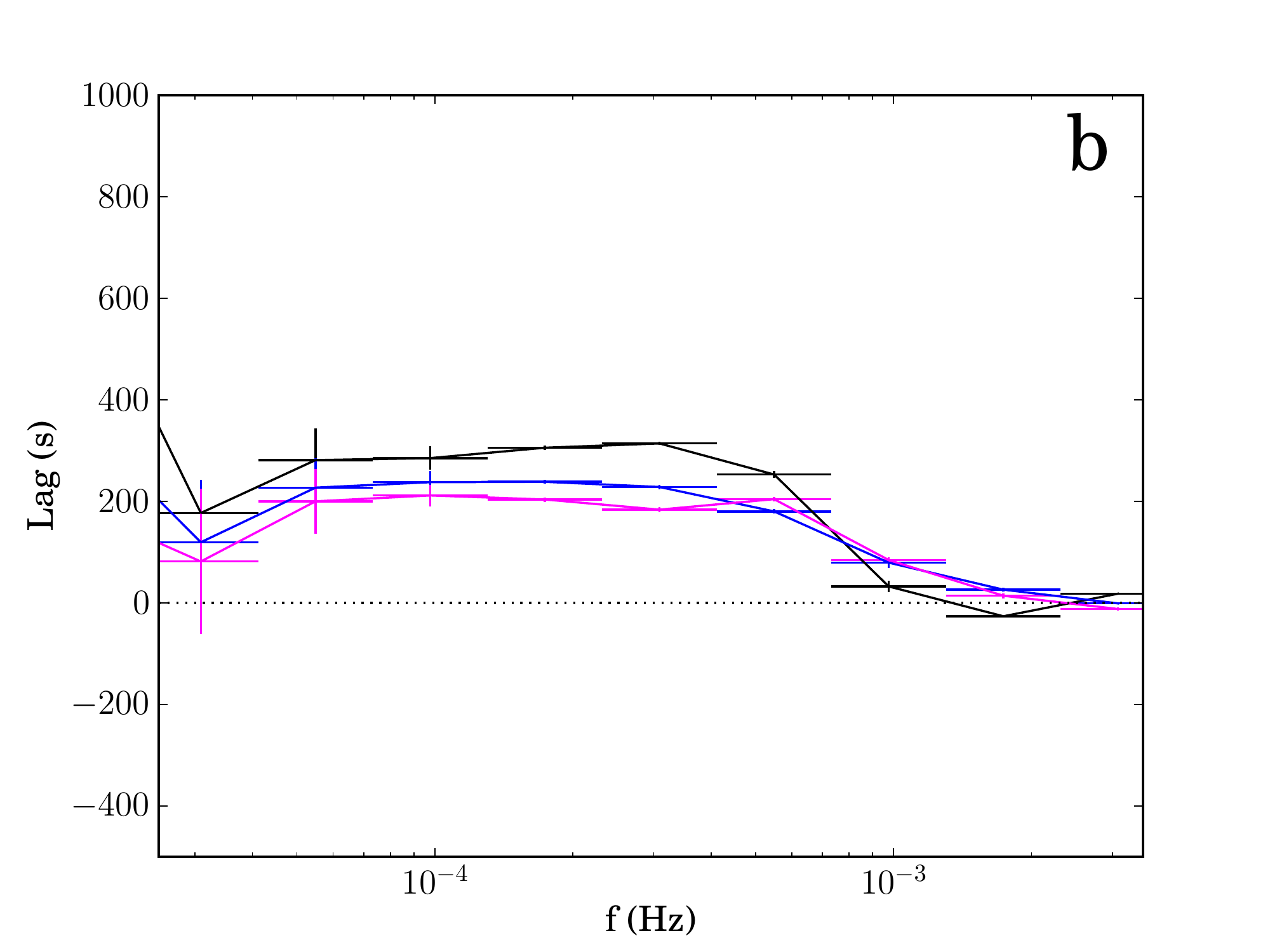} \\
\end{tabular}
\caption{Effect of increasing cloud radius on a). power spectrum and b). lag-frequency spectrum, for $R_{cl}=5$ (black), $10$ (blue) and $15R_g$ (magenta). Solid lines show soft band ($0.3-1$keV) power spectra, dashed lines show hard band ($2-5$keV) power spectra.}
\label{fig5}
\end{figure*}

\begin{figure*} 
\centering
\begin{tabular}{l|r}
\leavevmode  
\includegraphics[width=8cm]{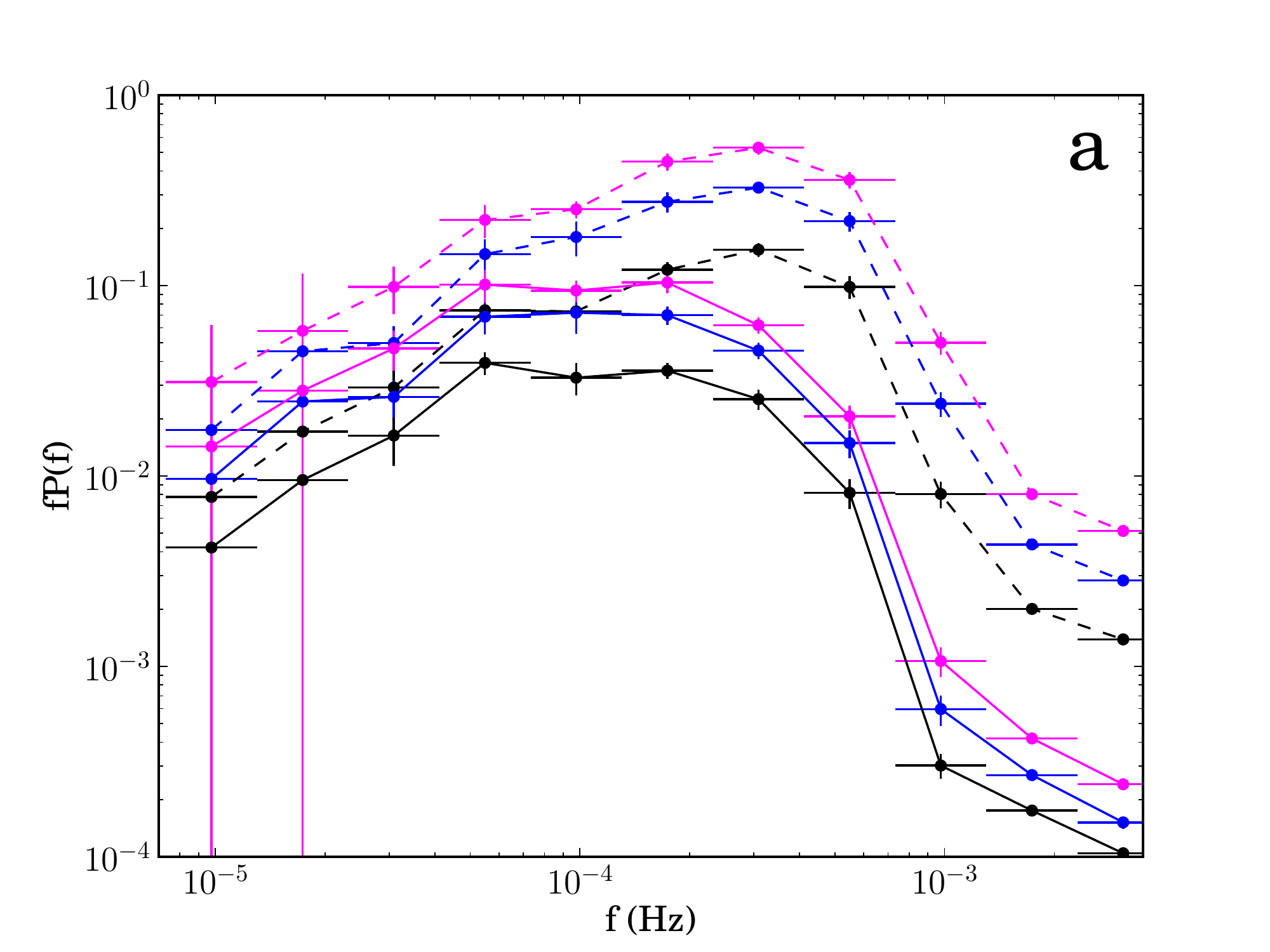} &
\includegraphics[width=8cm]{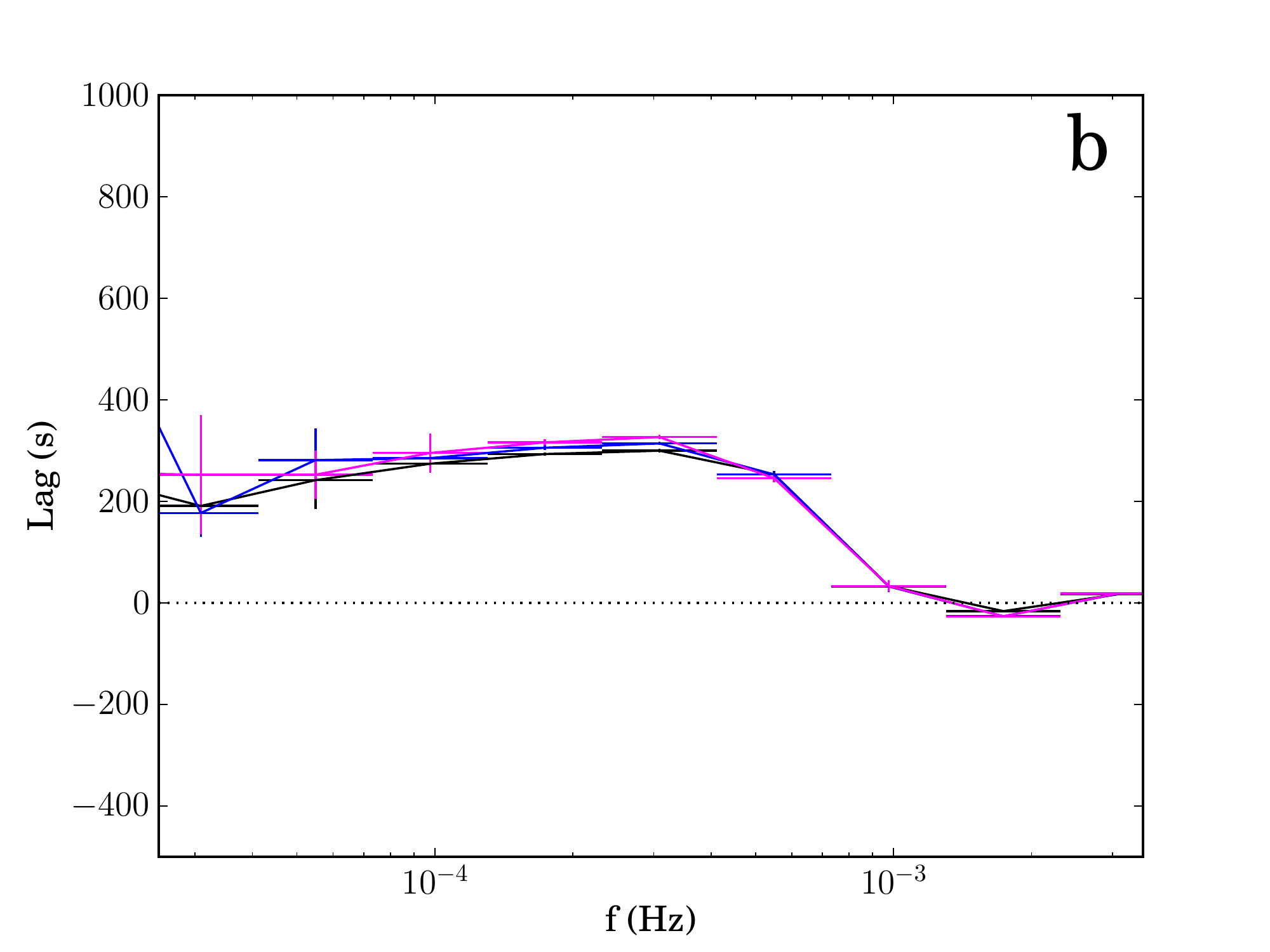} \\
\end{tabular}
\caption{Effect of increasing number of clouds on a). power spectrum and b). lag-frequency spectrum, for $n_{cl}\sim5\times10^{-4}$ (black), $10^{-3}$ (blue) and $1.5\times10^{-3}$ (magenta) clouds per second. Solid lines show soft band ($0.3-1$keV) power spectra, dashed lines show hard band ($2-5$keV) power spectra.}
\label{fig6}
\end{figure*}

\subsection{Effect of Transit Time}

Fig. 4 shows the effect of changing the cloud transit time. We increase
the transit time from $T_{tr}=5\times10^3$s (black) to
$1.5\times10^4$s (magenta). The hard band power spectra (dashed lines,
Fig. 4a) clearly show that as the transit time increases, the peak
frequency at which power is added to the light curves decreases. The
decrease in peak frequency of $\sim$ half an order of magnitude
roughly matches the threefold increase in transit time. The total
amount of power added to the light curves also increases by a similar
amount in both bands.

Fig. 4b shows how this affects the lag measured between the hard and
soft band light curves. For short transit times ($5\times10^3$s,
black), a short lag ($\sim150$s) is measured up to high frequencies
($10^{-3}$Hz). As the transit time increases, the maximum frequency at
which a lag is measured decreases. This is because occultations with a
longer transit time cannot add high frequency power to the light
curves (as shown by the power spectra in Fig. 4a). The absolute value
of the lag also increases, since for longer transit times the clouds
spend longer occulting the outer soft components before they cross and
occult the central corona. The measured lag drops from $\sim550$ -
$150$s roughly matching the decrease in transit time of a third from
$1.5\times10^4$ to $5\times10^3$s. 

We note that increasing black hole mass or increasing the size scales
of the individual components has a similar effect on the power
spectrum and variability as increasing the transit time. 

\subsection{Effect of Cloud Radius}

We now fix the transit time at $10^{4}$s and investigate the effect of changing the cloud radius. Fig. 5 shows the resulting power spectra and lag-frequency spectra for $R_{cl}=5$, $10$ and $15R_g$ (black, blue and magenta lines, respectively). 

Increasing the cloud radius increases the amount of power in the light curves. The soft band shows the biggest increase (Fig. 5a, solid lines), with the amount of power at $10^{-4}$Hz increasing by nearly one and a half orders of magnitude. The effect is much less in the corona dominated hard band (just under an order of magnitude), because the corona is much smaller, so that it is already completely obscured by a small cloud of $5R_g$. Increasing the cloud radius only prolongs the length of time it is obscured. By contrast the much larger disc is never completely obscured by a $5R_g$ cloud. Increasing the cloud radius therefore increases the area of the disc that experiences obscuration and hence adds more power to the soft band light curve. 

The more noticeable change to the hard band light curve is that the frequency at which the hard band power peaks decreases as the cloud radius increases (Fig. 5a, dashed lines). This is because, for a larger cloud radius (and fixed transit time), the time taken between covering and uncovering the corona increases. Consequently the transit cannot add as much high frequency power to the light curve. The peak in power drops from $\sim3\times10^{-4}$ to $2\times10^{-4}$, as more power is added at low frequencies and lost at high frequencies. 

Fig. 5b shows the lag as a function of frequency between the hard and soft band light curves. The lag measured actually slightly decreases as cloud radius increases. This is because the larger the cloud the more time it spends obscuring hard and soft components simultaneously. This results in very broad, very similar flux drops in both the hard and soft bands. In contrast, the strongest soft leads are seen when the cloud is small enough to obscure the blue wing of the disc and soft excess and then the corona in turn. This results in much narrower flux drops in the hard and soft bands, where the skew due to Doppler boosting (which causes the soft lead) is much more prominent.

\subsection{Effect of Cloud Number Density}

Fig. 6 shows the effect of increasing the number of occulting clouds. We fix the cloud radius and transit time at $5R_g$ and $10^{4}$s and increase the number density of clouds from $n_{cl}\sim5\times10^{-4}$ (black) to $1.5\times10^{-3} s^{-1}$ (magenta). Our total simulation time is $1.024$Ms, in practice this corresponds to increasing the total number of occulting clouds from 500 to 1500 clouds, each of which is assigned a random start time for its transit. 

Fig. 6a shows that increasing the number of clouds increases the power spectrum normalisation without affecting its shape. This is because more occultations simply add more power to the light curve. Both the hard and soft bands are affected equally. Fig. 6b shows that increasing the number of occultations has no effect on the lag measured between the hard and soft band light curves. The value of the lag is determined primarily by the transit time, with a weak dependence on cloud radius. 

\begin{figure} 
\centering
\begin{tabular}{l}
\leavevmode  
\includegraphics[width=8cm]{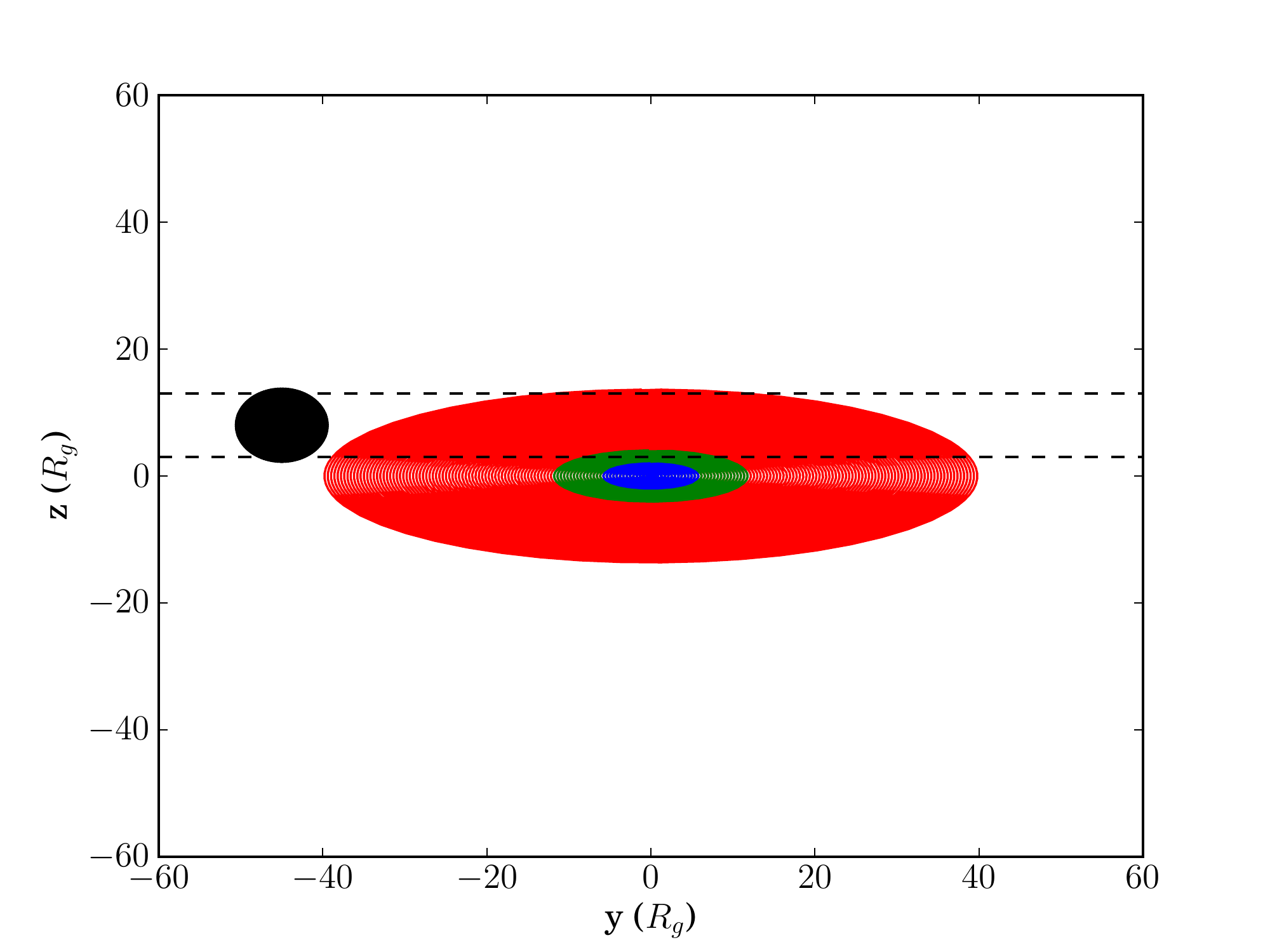} \\
\end{tabular}
\caption{Dashed lines show path of cloud across accretion flow for a high transit latitude ($z_{cl}=8$), with disc (red), soft excess (green), corona (blue), obscuring cloud (black) and accretion flow viewed at $i=70^\circ$.}
\label{fig7}
\end{figure}

\begin{figure*} 
\centering
\begin{tabular}{l|r}
\leavevmode  
\includegraphics[width=8cm]{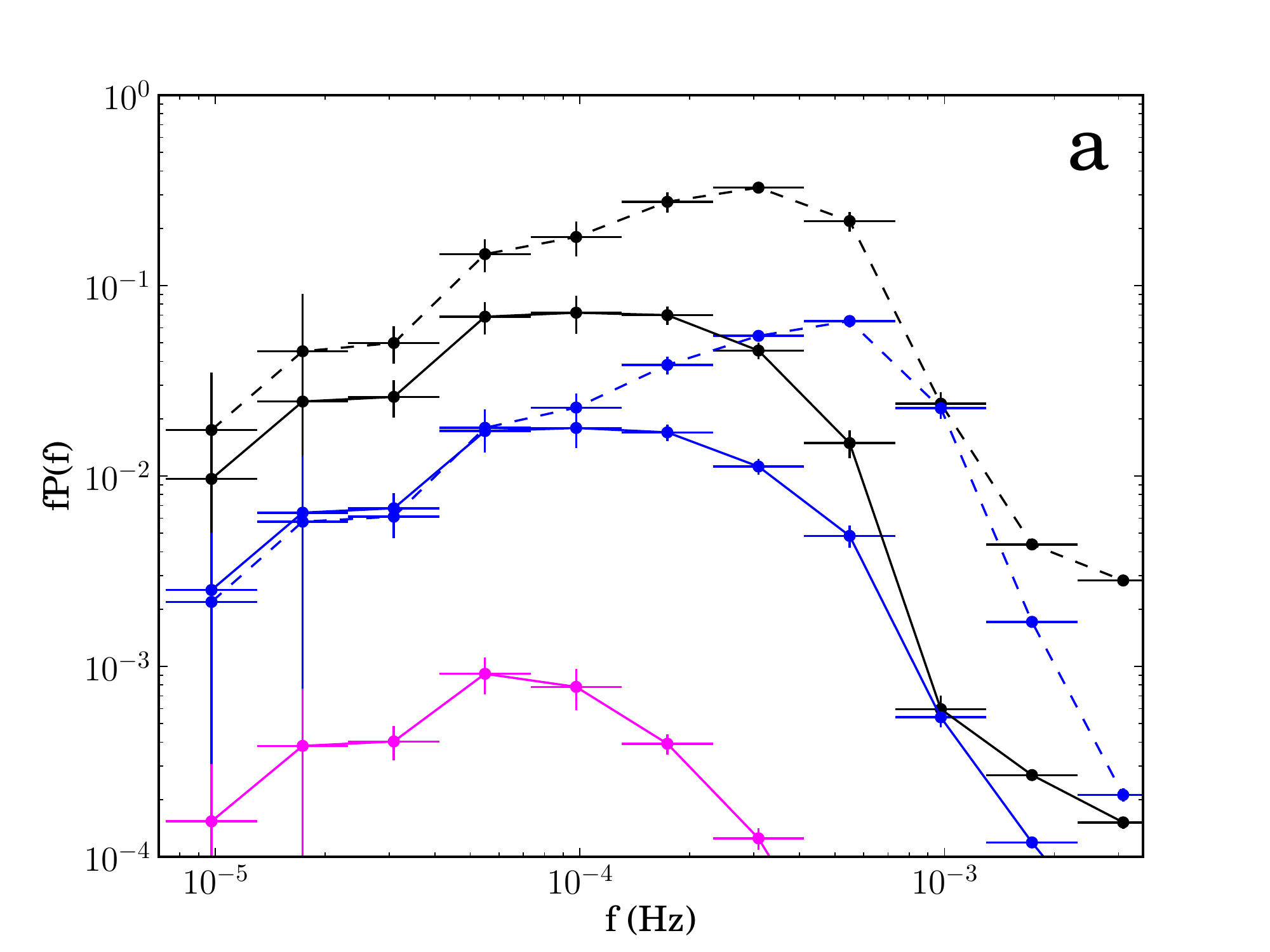} &
\includegraphics[width=8cm]{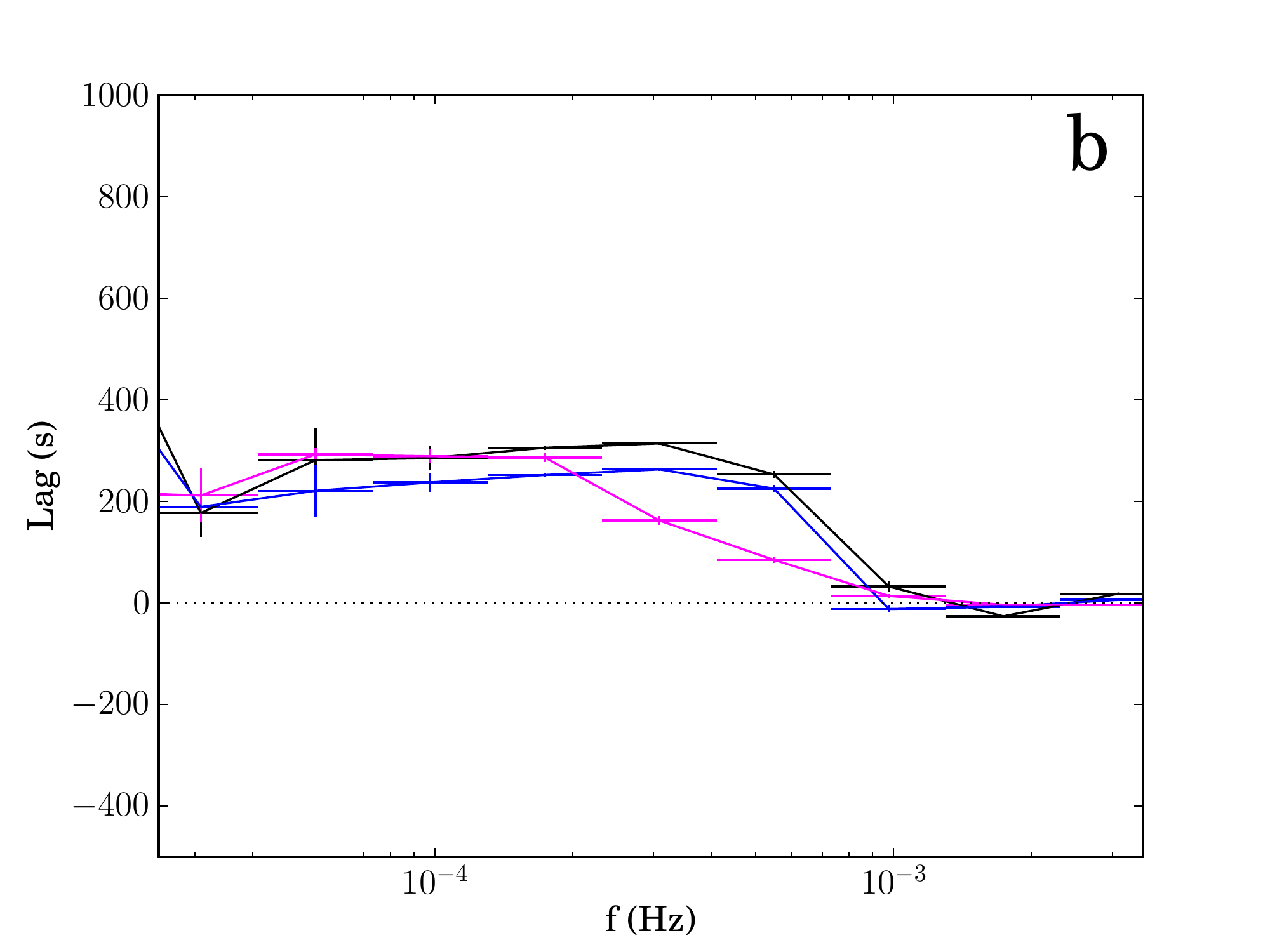} \\
\end{tabular}
\caption{Effect of increasing transit latitude on a). power spectrum and b). lag-frequency spectrum, for $z_{cl}=0$ (black), $5$ (blue) and $8R_g$ (magenta), where $z_{cl}$ defines the apparent latitude of the centre of the clouds as shown in Fig. 1a. Solid lines show soft band ($0.3-1$keV) power spectra, dashed lines show hard band ($2-5$keV) power spectra.}
\label{fig8}
\end{figure*}

\begin{figure*} 
\centering
\begin{tabular}{l|c|r}
\leavevmode  
\includegraphics[trim = 0.3in 0.0in 0.5in 0.0in,clip,scale=0.3]{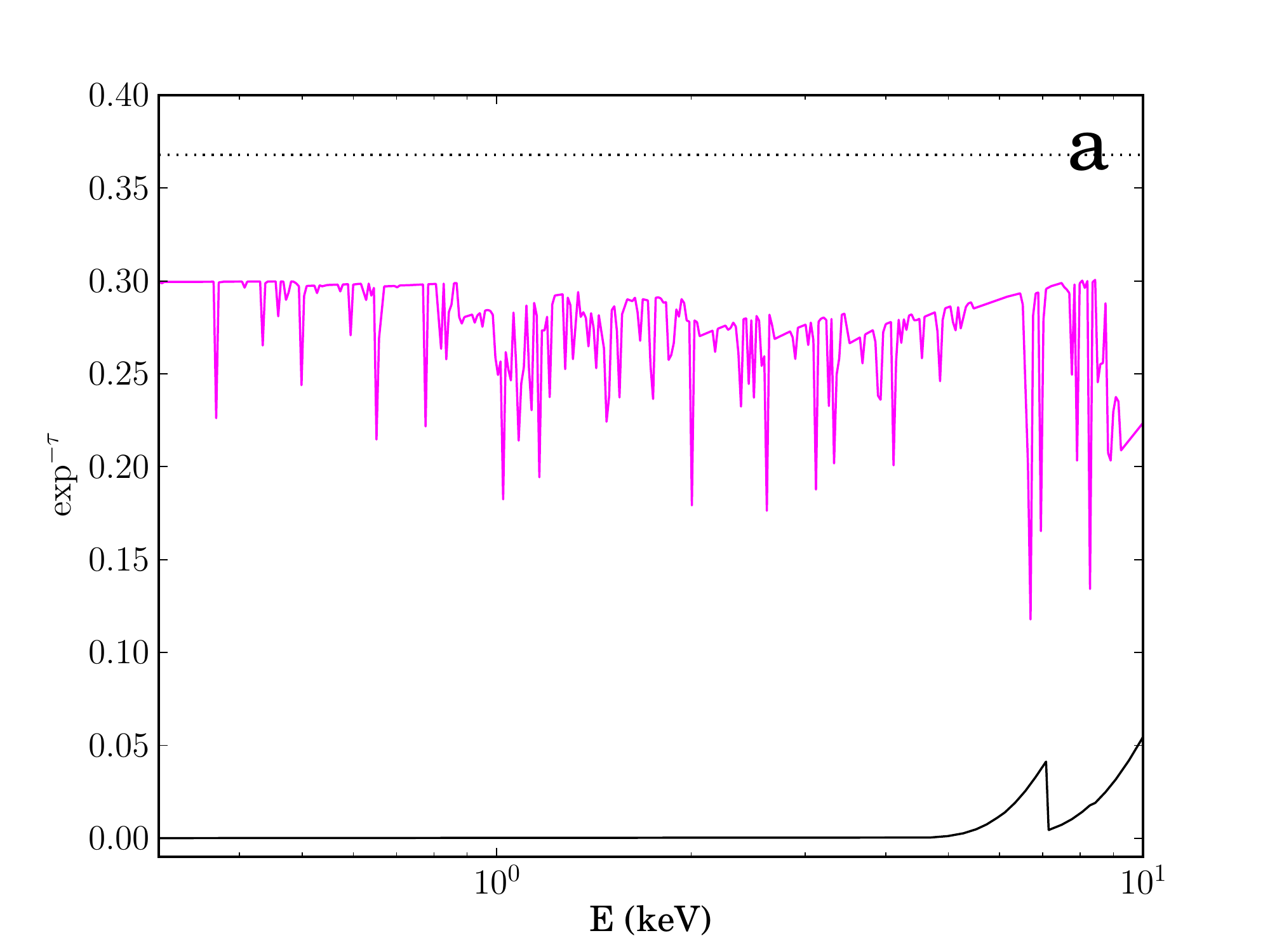} &
\includegraphics[trim = 0.25in 0.0in 0.5in 0.0in,clip,scale=0.3]{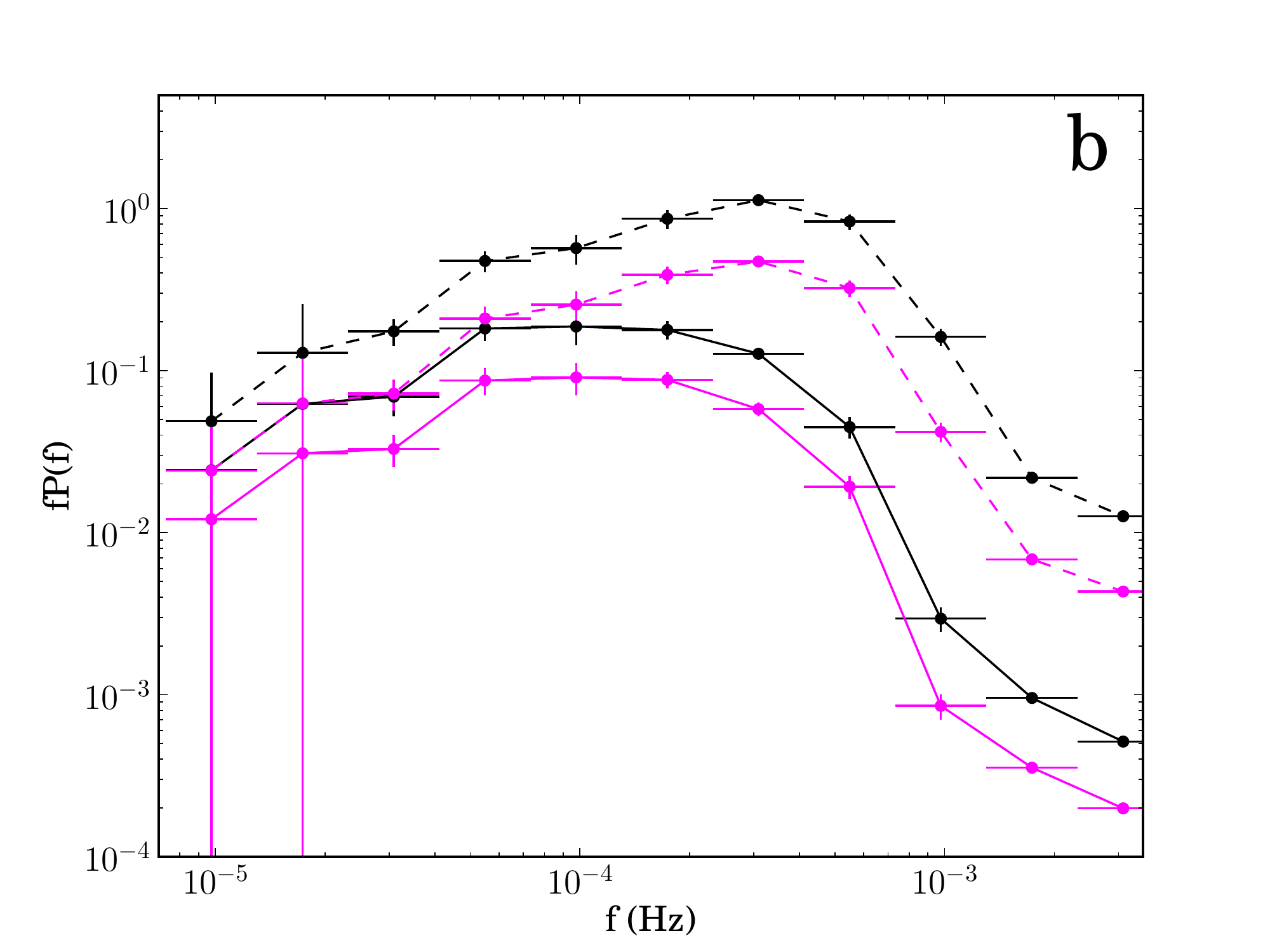} &
\includegraphics[trim = 0.15in 0.0in 0.5in 0.0in,clip,scale=0.3]{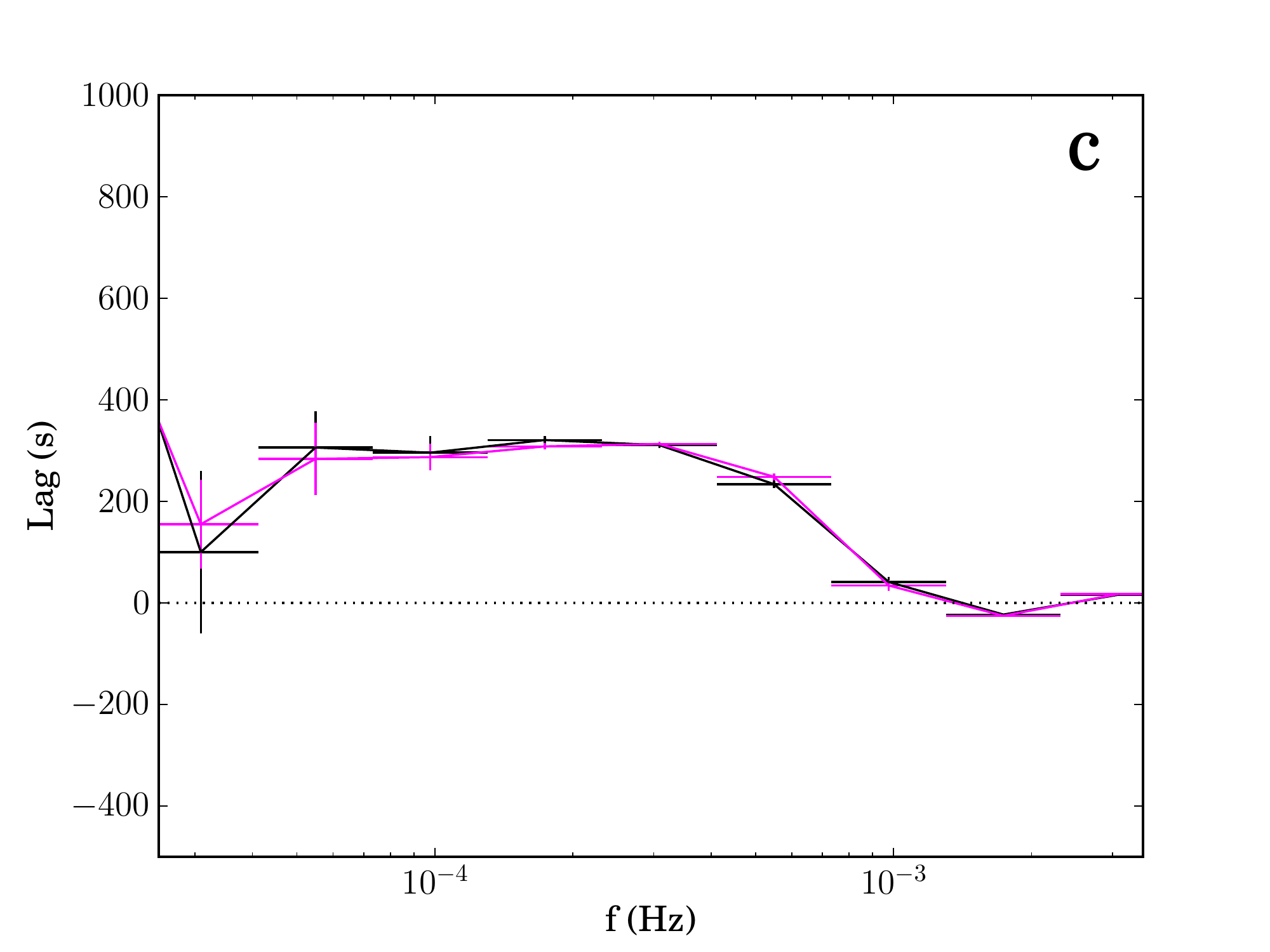} \\
\end{tabular}
\caption{Effect of changing cloud ionisation on a). transmission
spectrum, b). power spectrum and c). lag-frequency spectrum, for two
cloud ionisation states: $\xi=0$ (black) and $\log\xi=4$
(magenta). Solid lines show soft band ($0.3-1$keV) power spectra,
dashed lines show hard band ($2-5$keV) power spectra. Dotted line
shows transmitted flux level for constant $e^{-\tau}$, $\tau=1$.}
\label{fig9}
\end{figure*}

\subsection{Effect of Transit Latitude}

In Fig. 8 we show the effect of increasing the latitude of the cloud
path so that it no longer aligns exactly with the
black hole. In
all three cases we fix the cloud radius and transit time to $5R_g$ and
$10^{4}$s and the number density of clouds to $10^{-4}s^{-1}$. We
increase the apparent latitude of the cloud center from $z_{cl}=0$ to
$z_{cl}=8$.
Fig. 7 shows as an illustration the path taken by a cloud
transiting at $z_{cl}=8$, where the dashed lines bound the region
experiencing obscuration.

Fig. 8a shows  the effect on the hard and soft  band power spectra. As
the latitude increases the cloud  only obscures the 'back half' of the
accretion flow ($z>0$ in Fig.  7). Consequently the power in the light
curves decreases. By the time the latitude has increased to $8R_g$ the
cloud no longer obscures the  central corona, hence there is almost no
power in the  hard band (hence no dashed magenta  line), since we have
fixed the underlying accretion flow to have constant
flux. Consequently high latitude transits that do not occult the
corona add power to the soft band that is uncorrelated with the hard
band.

The soft band power drops from $\sim7\times10^{-2}$ for a central
transit (solid black line) to $\sim10^{-3}$ at the highest latitude
(solid magenta line). Comparison of the black and magenta solid lines
shows that, not only has total power been lost, but power has
preferentially been lost at high frequencies. Below $5\times10^{-5}$Hz
the two soft band power spectra show the same shape, whilst above
$5\times10^{-5}$Hz the high latitude power spectrum shows a cut
off, with the power dropping off sharply above $10^{-4}$Hz. This is
because the transiting cloud just clips the
very edge of the disc and soft excess. These regions of the flow are
travelling nearly perpendicular to the line of sight so experience
very little Doppler boosting/deboosting. They therefore carry only a
moderate fraction of the total soft band flux. The shortest, sharpest
dips in the soft band light curve arise through the cloud occulting
the innermost parts of the disc/soft excess, which are centrally
concentrated and strongly Doppler boosted. These add the highest
frequency components to the soft band light curve. Occultations of the
outer parts of the flow result in slower more gradual flux drops and
hence add low frequency power. The coronal power law also contributes
some flux to the soft band, so with no coronal occultations that
removes its additional source of high frequency power.

Fig. 8b shows the effect of increasing the transit latitude on the
lag measured between the hard and soft bands. As high frequency power
is lost from both the hard and soft bands, the measured lag begins to tend to zero at lower frequencies ($\sim2\times10^{-4}$Hz for the high latitude magenta spectrum, compared to $7\times10^{-4}$Hz for the central transit back spectrum). The lag tends more gradually to zero in the case of the high latitude transit. This is again because the highest frequency components come from the shortest, sharpest flux drops which arise from occulting the brightest central regions. For a high latitude transit this is when the cloud just clips the small part of the soft excess that appears at high latitude (i.e. large $z$ and $y=0$ in Fig. 7) at the midpoint of the transit. Because the coronal flux is not occulted, occultation of the soft excess is the only source of variability in the hard band. It is also the only source of high frequency variability in the soft band. When the source of variability is the same in both bands, there can be no lag between them. Hence as frequency increases above $\sim2\times10^{-4}$Hz the lag tends gradually to zero as the only source of variability becomes occultation of the soft excess in both bands.

\subsection{Effect of Cloud Ionisation}

So far we have approximated the absorption of the cloud as
$e^{-\tau}$, where the electron scattering optical depth, $\tau=1$
(equivalent to a pure hydrogen column of $1.5\times
10^{24}$~cm$^{-2}$, which is $1.25\times 10^{24}$~cm$^{-2}$ for solar
abundance material), is a constant with energy. However this is only
appropriate for completely ionised material. In general, the optical
depth of the cloud is a function of energy, depending on the
ionisation state of the cloud.

In Fig. 9a we show our original model (dotted black line) compared to
transmission spectra for $N_H=1.5\times 10^{24} cm^{-2}$ at $\log \xi=4$
(magenta line) and at $\xi=0$ (solid black line) The $\log \xi=4$
transmission spectrum is calculated using the {\sc{xspec}} model
{\sc{zxipcf}}. This assumes a turbulent velocity of 200~km/s, so the line
strength can be enhanced for the same column density of material for higher
velocities. The neutral spectrum is calculated using
{\sc{phabs}}. As before, we fix $R_{cl}=5R_g$, $T_{tr}=10^{4}$s,
$n_{cl}=10^{-4}s^{-1}$ and $z_{cl}=0$. Having fixed both the optical
depth and cloud radius, this constrains the cloud density, since
$n=\tau/(2R_{cl}\sigma_T)\sim10^{11}cm^{-3}$. The ionisation state of
the clouds is related to their density and the X-ray luminosity as
$\xi=L_X/(nD^2)$, where $D$ is the distance of the cloud from the
central X-ray source. For $L_X=10^{42}erg s^{-1}$ and $D=20R_g$,
$\xi\sim10^4$. 

Fig. 9b shows that increasing the ionisation state of the cloud reduces
the amount of power added to the hard and soft band light curves. The
difference is roughly half an order of magnitude in both bands, since
the change in opacity is roughly the same for both bands. For a
neutral cloud, the transmitted flux below $5$keV is zero.
In contrast, when the cloud is highly
ionised there is very little absorption left, so the
fraction of
transmitted flux rises to $e^{-\tau}\sim0.3$. This results in shallower flux drops and hence less power in the hard and soft band light
curves.

Fig. 9c shows that changing the ionisation state of the clouds has no
effect on the lag measured between the hard and soft energy bands. The
lag depends on the motion of the cloud with time, not on the relative
amounts of power carried by the hard and soft bands.

\begin{figure*} 
\centering
\begin{tabular}{l|r}
\leavevmode  
\includegraphics[width=8cm]{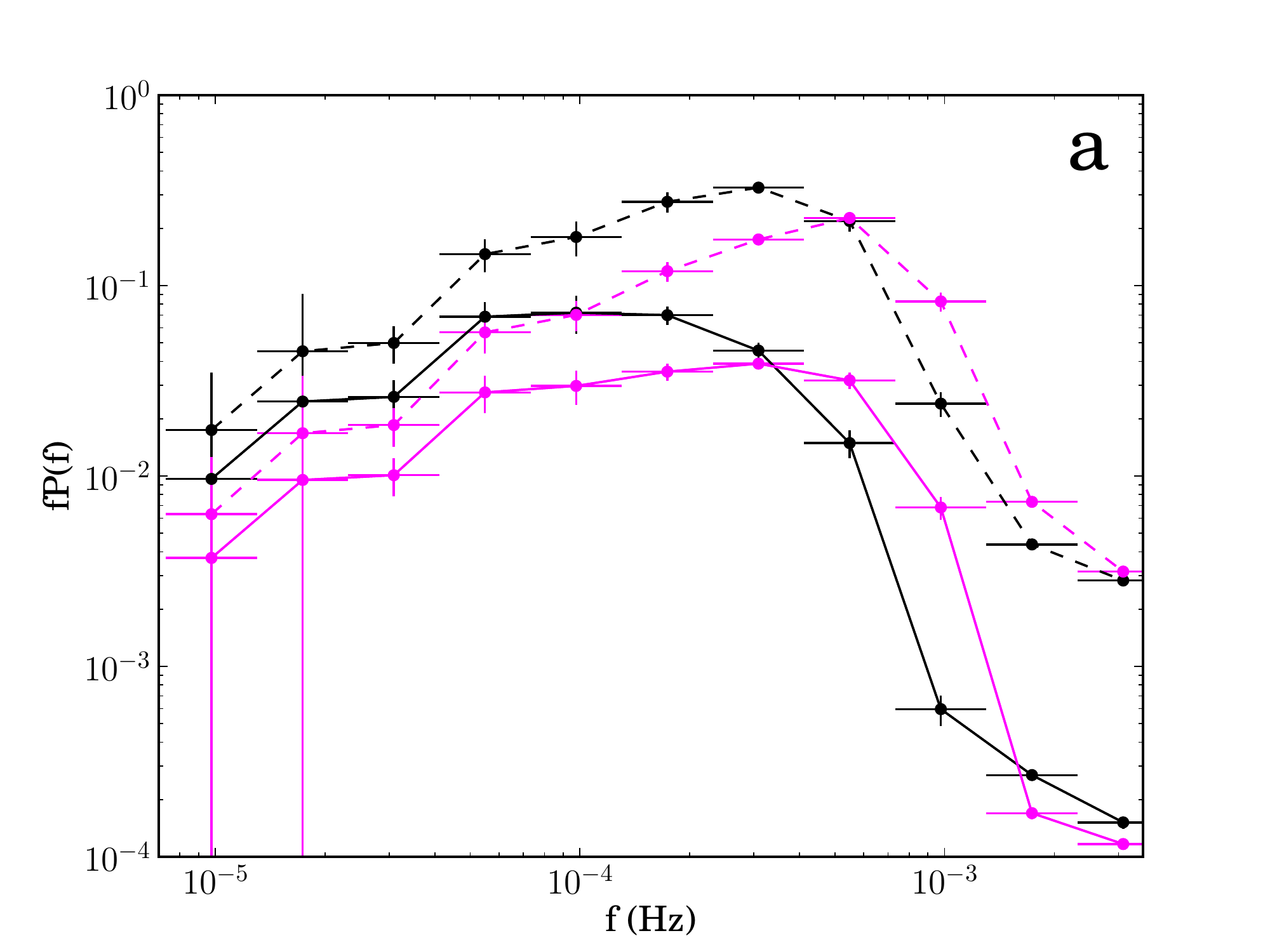} &
\includegraphics[width=8cm]{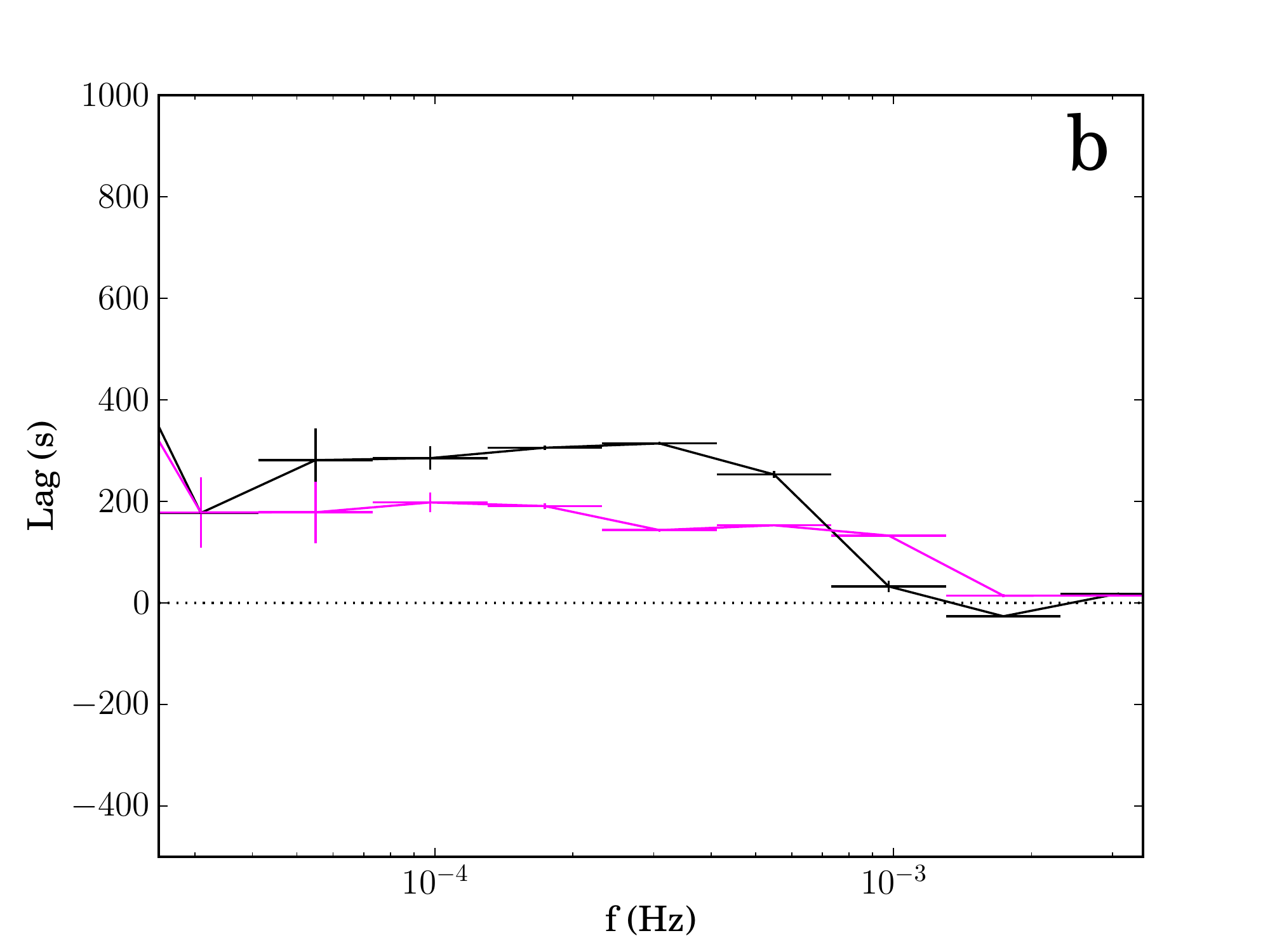} \\
\end{tabular}
\caption{Change in a). power spectrum and b). lag-frequency spectrum
when circular motion of clouds is taken into account (magenta),
compared with linear occultations (black). Solid lines show soft band
($0.3-1$keV) power spectra, dashed lines show hard band ($2-5$keV)
power spectra.}
\label{fig10}
\end{figure*}

\section{Circular Occultation}

So far we have modelled linear occultations, where the apparent
velocity of the clouds remains constant during the transit. However
the clouds should be rotating with the accretion flow, in which case
their apparent velocity during the transit will vary as a cosine
function. During the middle of the transit the component of the
cloud's velocity perpendicular to the line of sight is greatest and
the cloud appears to move faster. At the beginning and end of the
transit the cloud is moving towards/away from the observer, the
component of its velocity perpendicular to the line of sight is small
and its apparent velocity is much slower. If the orbital radius of the
clouds is much larger than the radius of the region being occulted
then linear occultation is a reasonable approximation. As the orbital
radius of the clouds becomes similar to the occulting region size the
effect becomes more important. A transit time of $10^4$s corresponds
to a Keplerian orbital velocity at $\sim20R_g$ for a $10^7M_\odot$ BH,
implying this effect should be taken into account.

Fig. 10 shows the effect on the power spectra and lag-frequency spectrum. We fix $R_{cl}=5R_g$, $T_{tr}=10^{4}$s, $n_{cl}=10^{-4}s^{-1}$ and $z_{cl}=0$ and show the result of linear occultations in black and accounting for circular motion in magenta. In the circular case the cloud moves faster while it is occulting the brightest central region of the accretion flow. As a result the flux drops are narrower. This adds more power at higher frequencies, hence both the hard and soft band power spectra are shifted to slightly higher frequencies for the case of circular occultations (Fig. 10a). 

Similarly the lag-frequency spectrum extends to slightly higher
frequencies in the circular motion case, showing a non-zero lag up to
$10^{-3}$Hz compared to $7\times10^{-4}$Hz for linear occultations (Fig. 10b). The measured lag is also shorter ($\sim200$s compared to $300$s for linear occultations). Again this is a consequence of the cloud moving faster during the central part of the occultation. The lag predominantly arises from the delay between occulting soft excess and then coronal emission on the Doppler boosted side of the flow and when the cloud is moving faster the delay is shorter. 

\begin{figure*} 
\centering
\begin{tabular}{l|r}
\leavevmode  
\includegraphics[width=8cm]{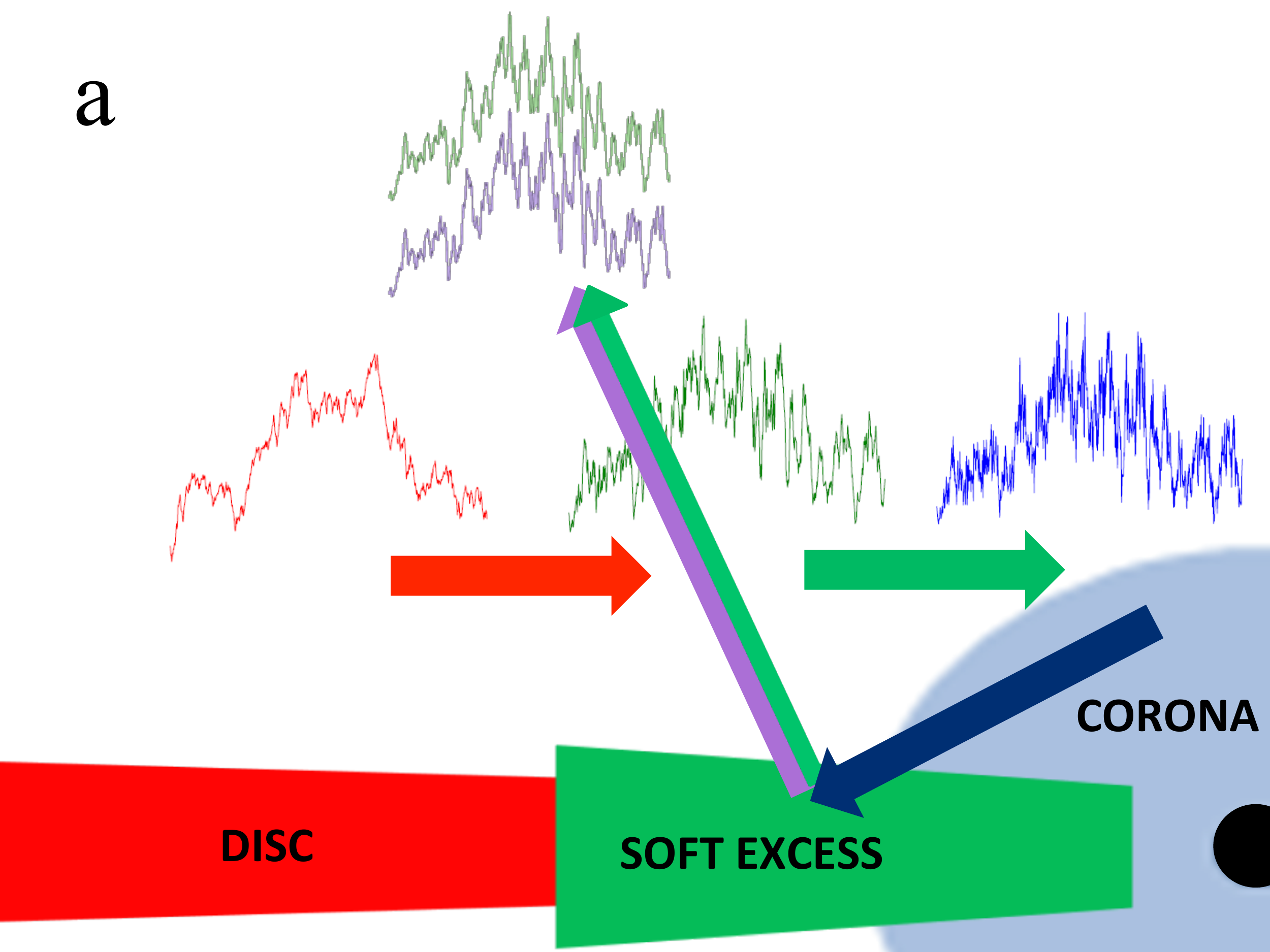} &
\includegraphics[width=8cm]{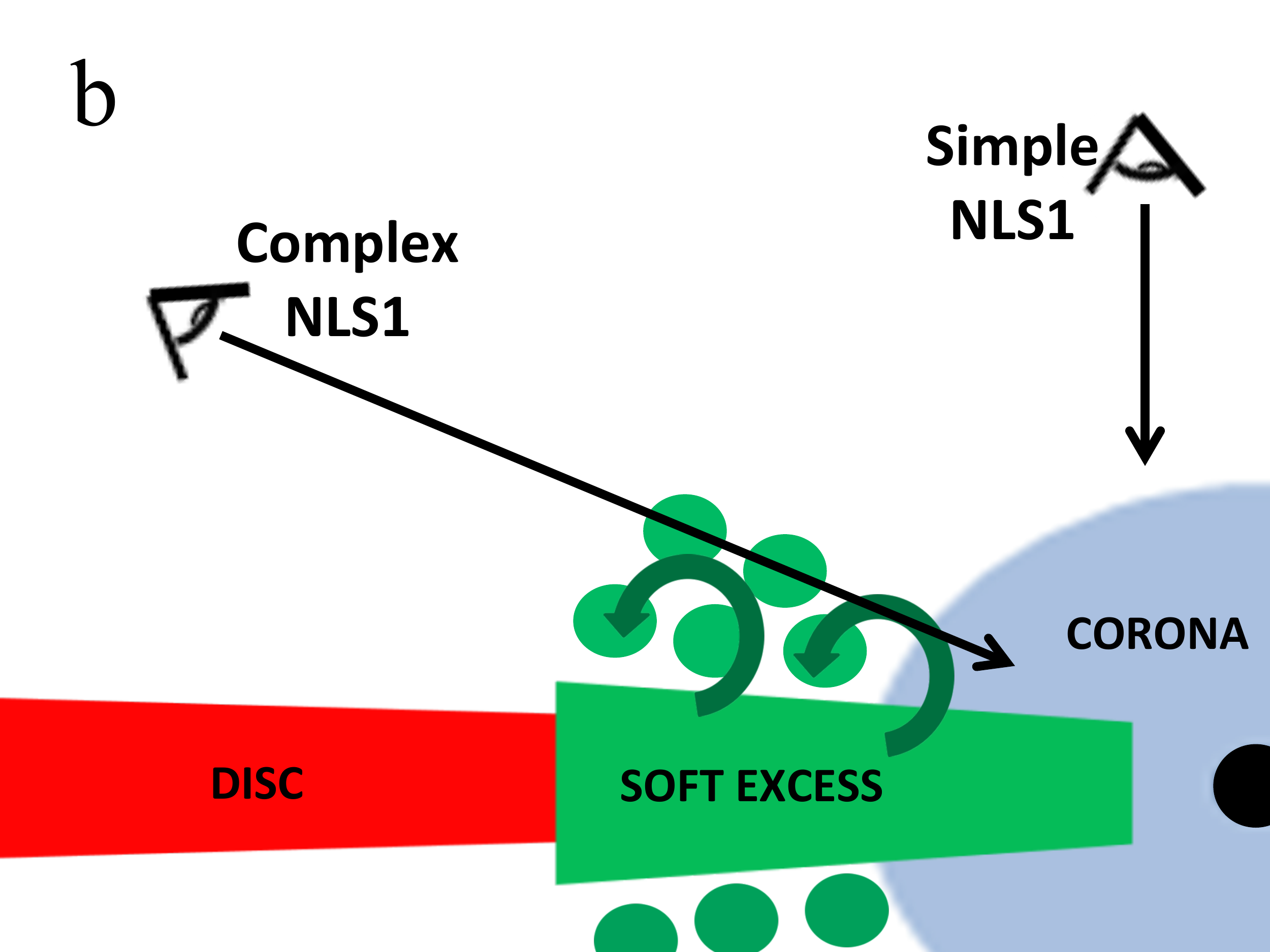} \\
\end{tabular}
\caption{a). Schematic of simple NLS1 model of GD14. Slow fluctuations propagate inwards from the outer components and are modulated by the faster fluctuations generated at smaller radii. The high energy coronal emission then reflects off and is reprocessed by the soft excess component. b). Scenario for transition from simple to complex NLS1 as a function of inclination, where the soft excess is a turbulent region of rotating clouds which partially obscures the line of site to the central regions in complex NLS1s. The clouds are then responsible for the bulk of the reflected/reprocessed emission, while partially obscuring the intrinsic emission.}
\label{fig11}
\end{figure*}

\begin{figure} 
\centering
\begin{tabular}{l}
\leavevmode  
\includegraphics[width=8cm]{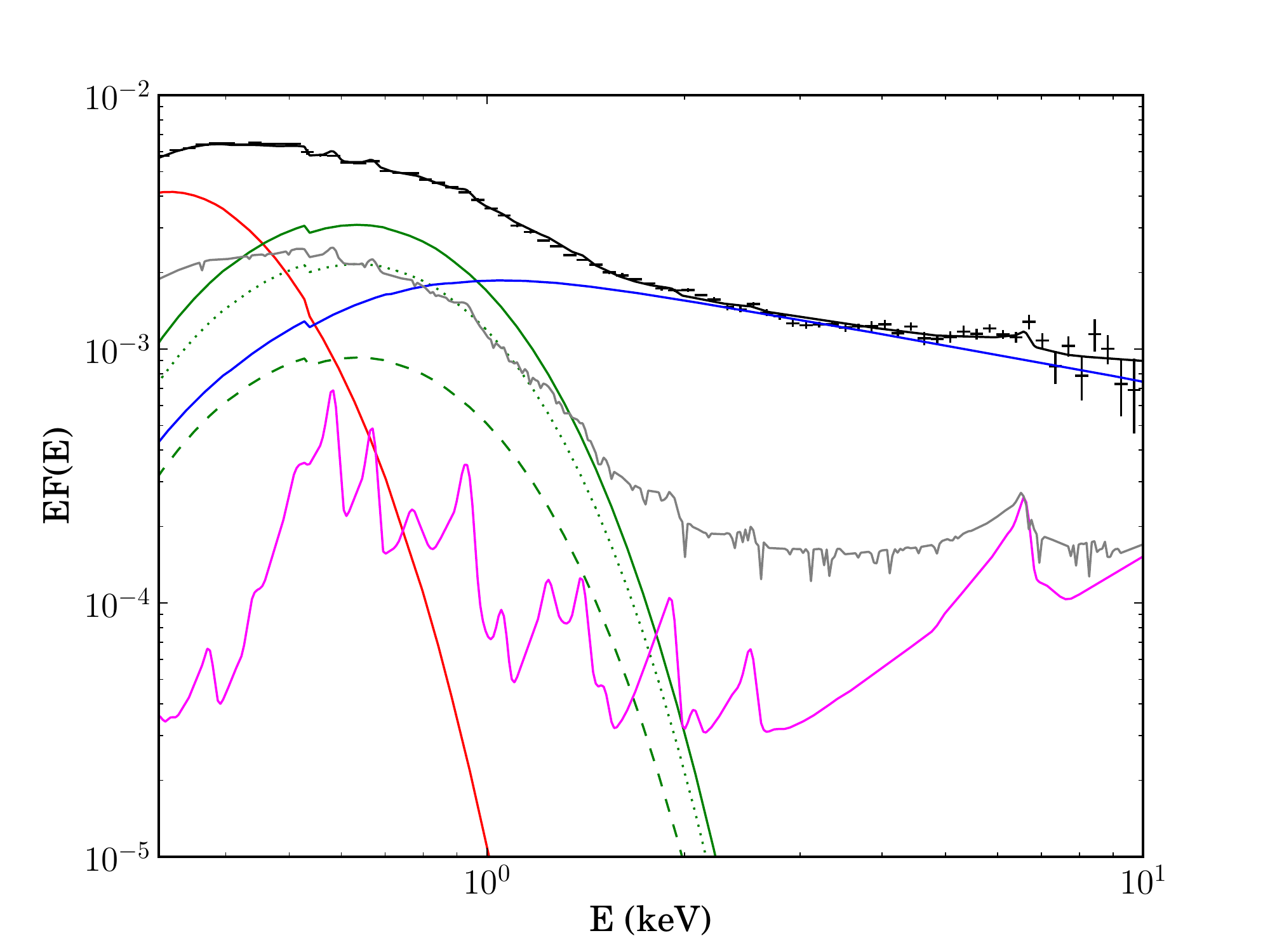}
\end{tabular}
\caption{Spectral decomposition for PG1244+026: disc (red), soft excess (green, with dashed line for intrinsic emission, dotted line for reprocessed emission, solid line for total), corona (blue), reflection (magenta), total (black). Data points show time averaged spectrum (OBS ID: 0675320101). Grey line shows an example of the spectrum after introducing absorption by intervening clouds.}
\label{fig12}
\end{figure}

\begin{figure*} 
\centering
\begin{tabular}{l|r}
\leavevmode  
\includegraphics[width=8cm]{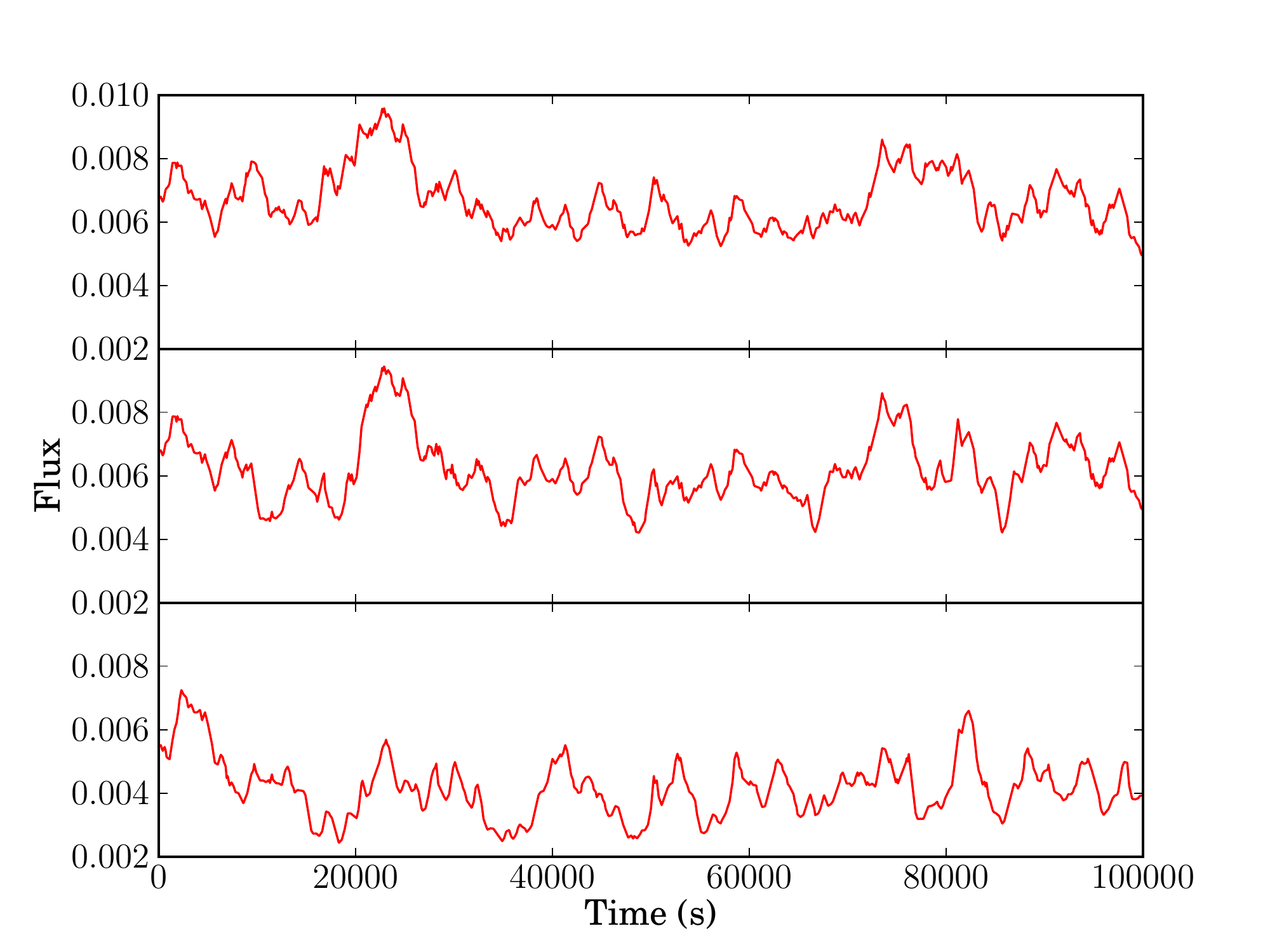} &
\includegraphics[width=8cm]{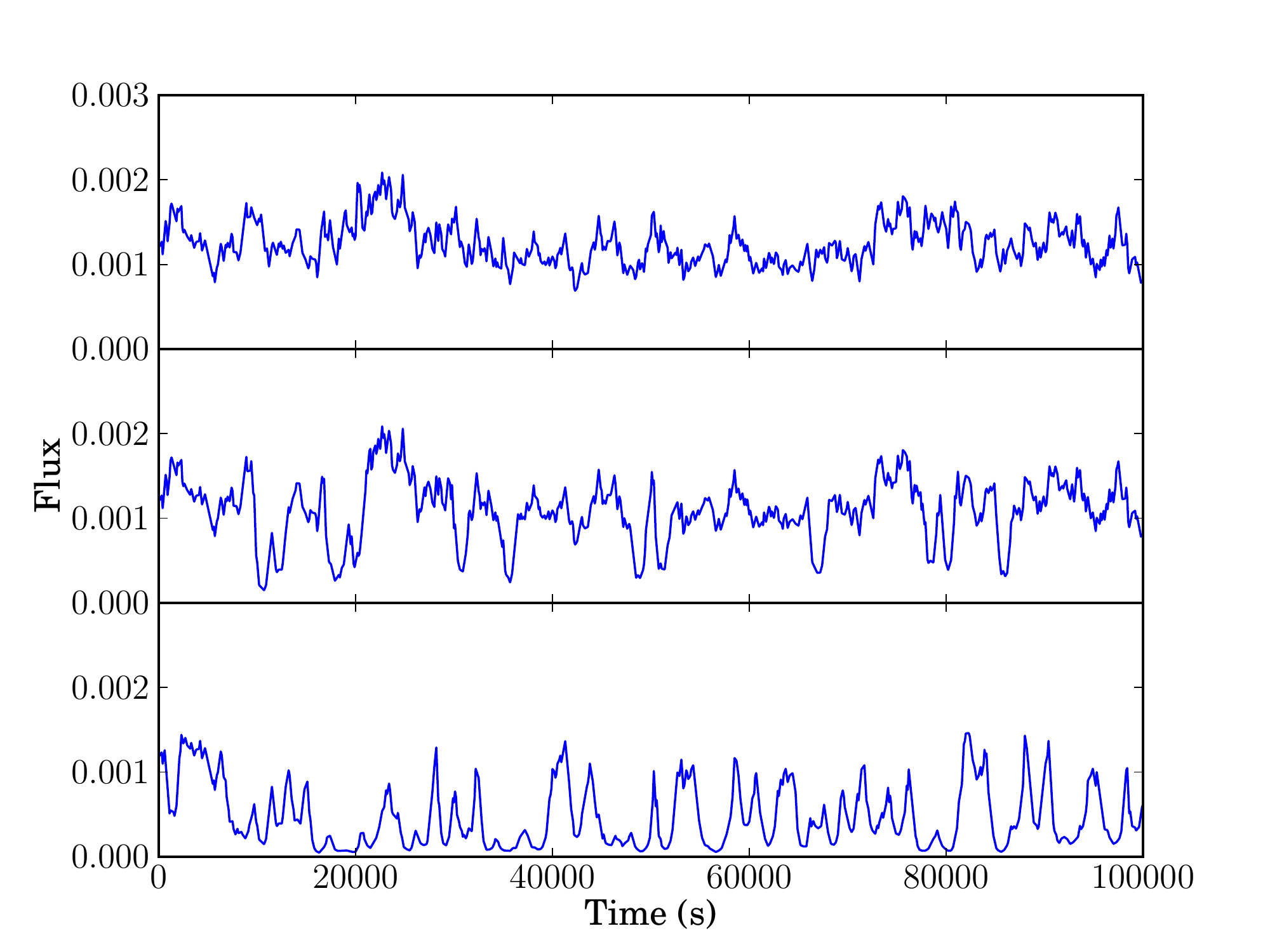} \\
\end{tabular}
\caption{Soft band ($0.3-1$keV, left panel) and hard band ($2-5$keV, right panel) light curves showing effect of adding obscuring clouds to the simple NLS1 model of GD14. Top panel: simple NLS1 model with no clouds. Middle panel: $n_{cl}=10^{-4}s^{-1}$. Bottom panel: $n_{cl}=10^{-3}s^{-1}$. For both cases we fix the cloud parameters to $R_{cl}=5R_g$, $T_{tr}=10^{4}$s, $z_{cl}=0$, $\log\xi=4$ and $N_H=1.5\times10^{24}cm^{-2}$ and assume the clouds are launched from $\sim20R_g$.}
\label{fig13}
\end{figure*}

\begin{figure*} 
\centering
\begin{tabular}{l|c|r}
\leavevmode  
\includegraphics[trim = 0.2in 0.0in 0.5in 0.0in,clip,scale=0.3]{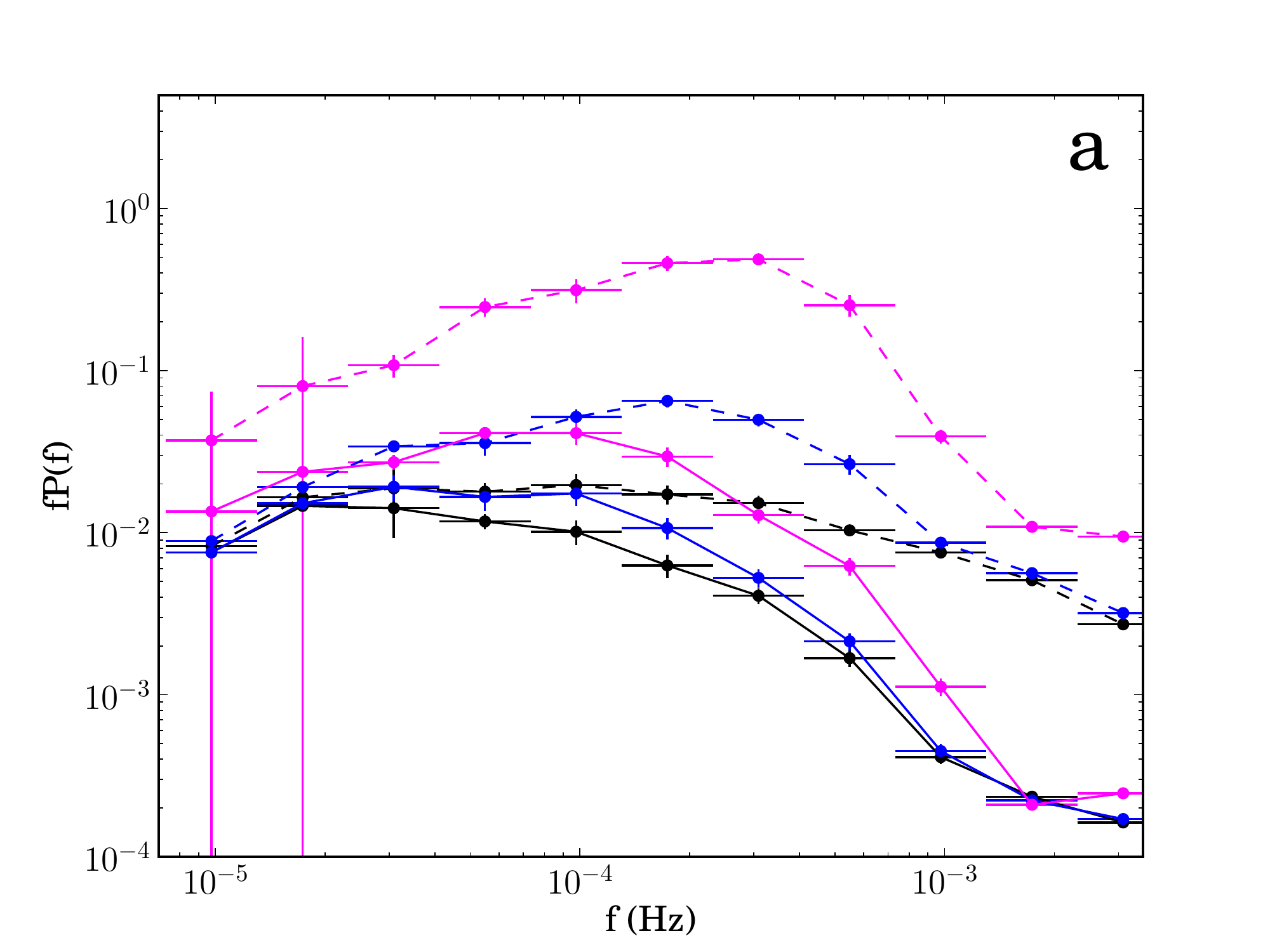} &
\includegraphics[trim = 0.15in 0.0in 0.5in 0.0in,clip,scale=0.3]{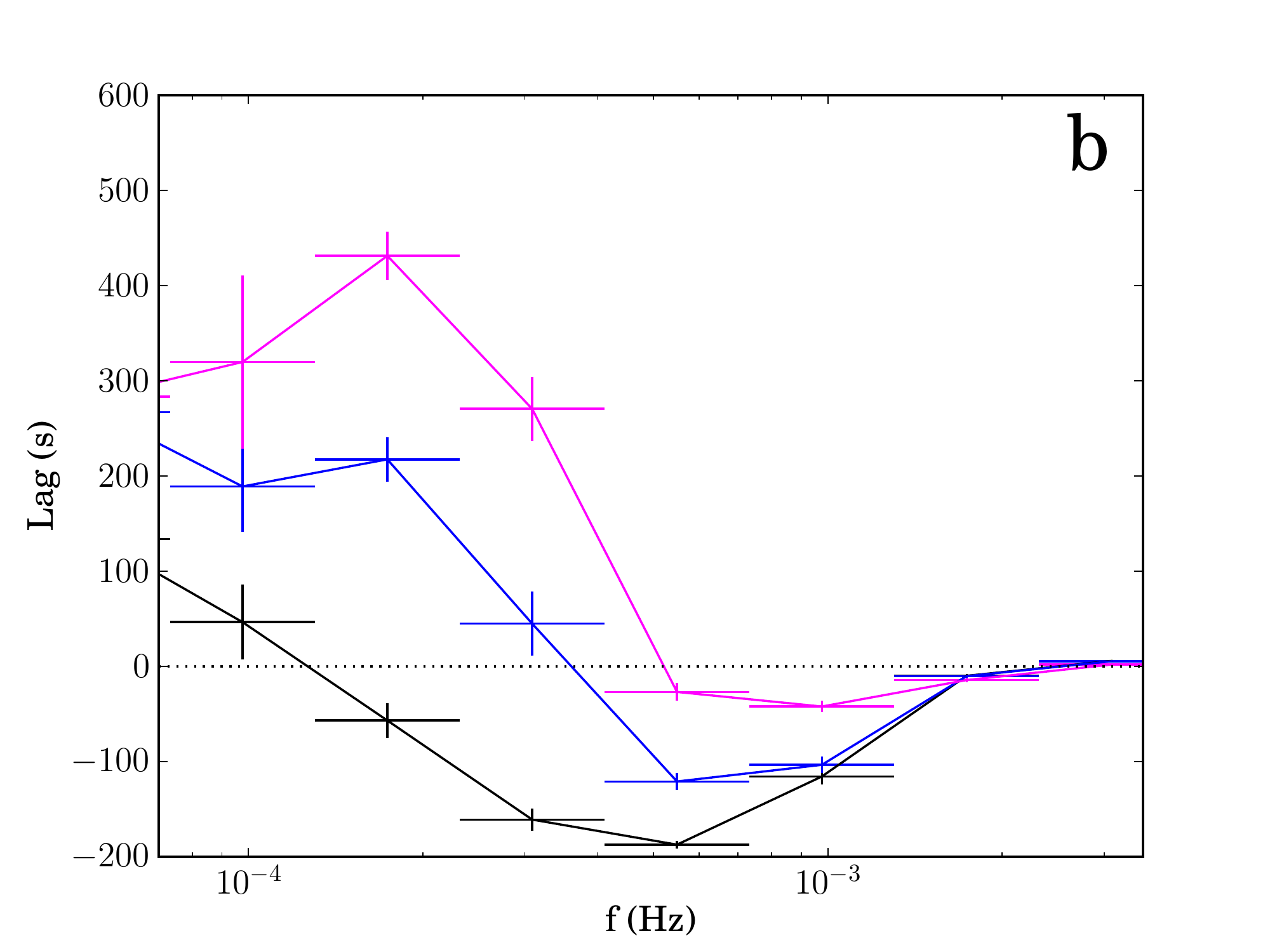} &
\includegraphics[trim = 0.3in 0.0in 0.5in 0.0in,clip,scale=0.3]{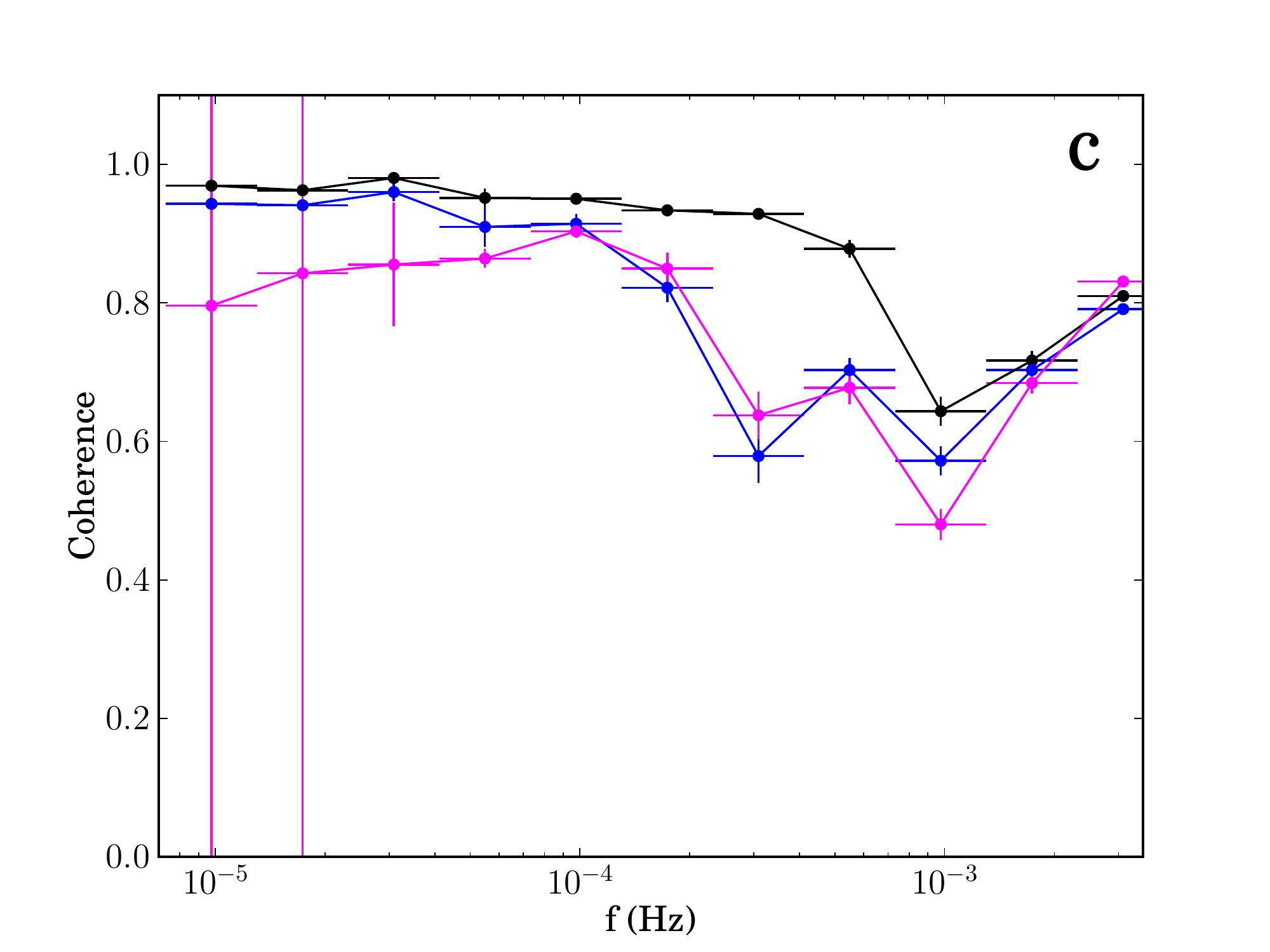} \\
\end{tabular}
\caption{Effect of adding occultations to the simple NLS1 model of GD14 on a). hard band ($2-5$keV, dashed lines) and soft band ($0.3-1$keV, solid lines) power spectra, b). lag frequency spectrum and c). coherence between hard and soft bands. Black lines show simple NLS1 model with no clouds, blue $n_{cl}=10^{-4}$ and magenta $n_{cl}=10^{-3}s^{-1}$, with cloud parameters as in Fig. 13.}
\label{fig14}
\end{figure*}

\section{Transition from Simple to Complex NLS1 by including occulting clouds}

We now investigate whether the addition of occultations can change the
timing properties of a simple NLS1 so that they appear more typical of
complex NLS1s. That is, can the effect of occultations reduce the
maximum measured reverberation lag from $\sim200$s to nearer $50$s and
shift it to higher frequencies?

In the previous sections we assumed constant flux from the underlying
accretion flow. We now replace this static model with the time
dependent model of GD14 shown in Fig. 11a.  The disc, which is at the
largest radii, generates the slowest fluctuations. These propagate
down to the soft excess, which is at smaller radii and generates its
own slightly faster fluctuations. The fluctuations in soft excess
emission therefore consist of the slow fluctuations from the disc,
delayed by some lag related to the propagation time, modulated by the
faster fluctuations generated in the soft excess. These fluctuations
then propagate down to the corona, which generates even faster
fluctuations. The hard coronal emission therefore shows fluctuations
on a whole range of timescales, as it responds to mass accretion rate
fluctuations propagating down from all radii. A fraction of these
central hard X-rays will illuminate the cooler soft excess and disc
components. Some of this illuminating flux will be reflected, the rest
will thermalise and be reprocessed. Fig. 12 shows our spectral
decomposition now including these reflected and reprocessed
components. These come from a fit to the time averaged spectrum of the
simple NLS1 PG1244+026 (OBS ID: 0675320101, see GD14), shown in black
data points. For simplicity we assume all reflection/reprocessing
occurs on the soft excess ($6-12R_g$). Thus the fluctuations in the
reflected/reprocessed emission follow the coronal fluctuations (Fig. 
11a), except for the very fastest fluctuations which are smoothed out
by the range of light travel time delays. Hence the reflected and
reprocessed fluctuations are a lagged and smoothed version of the hard
coronal fluctuations. The soft excess therefore consists of intrinsic
emission from the accretion flow (dashed green line, Fig. 12), which
varies slowly due to intrinsic mass accretion rate fluctuations in the
soft excess and those that have propagated inwards from the disc, and
reprocessed emission (dotted green line, Fig. 12) which follows the
faster coronal fluctuations. GD14 showed that this model can reproduce
all the observed timing properties of the simple NLS1 PG1244+026.

We use this model to describe the emission from the underlying
accretion flow as a function of time, and now add the effect of
occulting clouds. We fix the cloud parameters to $R_{cl}=5R_g$,
$T_{tr}=10^{4}$s and $z_{cl}=0$. A transit time of $10^{4}$s implies
an orbital radius of $20R_g$. Hence we reduce our transit radius from
$40R_g$ in Fig. 1a to $20R_g$ and take into account the circular motion
of the clouds. Since the clouds are launched so close to the central
X-ray source we allow them to be highly ionised and use the magenta
transmission spectrum shown in Fig. 9a ($\log\xi=4$,
$N_h=1.5\times10^{24}cm^{-2}$). $20R_g$ is consistent with the clouds being launched as
part of a failed Eddington (radiation pressure driven) wind from the
soft excess region as sketched in Fig. 11b. Radiation pressure lifts material from the accretion flow, which forms clumps as it rises (Takeuchi et al 2014). As soon as the optical depth of the clumps becomes $\tau>1$, some of the material is self-shielded from X-ray photons. The mass of the clump is still the same but the radiation pressure on it is now less. If the source is not strongly super-Eddington, the radiation pressure is not strong enough to expel the material so it falls back to the disc, resulting in a failed rather than outflowing wind. 
The whole turbulent large scale height region is
the source of the soft excess. Propagation of fluctuations occurs
through the disc regions which are the source of the intrinsic
emission, while the bulk of the reflected and reprocessed emission
comes from the turbulent clouds. We assume the turbulent velocity is
less than the orbital velocity ($v_{turb}<v_{Kepl}$) and that the clouds remain largely intact on the timescale of a single transit (although $v_{Kepl}$ may be sufficient to shred them on longer timescales, stripping off material before what remains falls back to the disc). For a
source at high inclination, these clouds will intercept the line of
sight to the central regions. As the clouds transit the line of sight,
we assume they obscure the intrinsic disc, intrinsic soft excess and coronal
emission. We do not obscure the reflected or reprocessed emission,
since we assume these are predominantly from the 
clouds.

\subsection{Fourier Timing Properties}

Fig. 13 shows the resulting soft and hard band light curves (left and
right respectively). The top panels show the original simple NLS1
model with no occultations. These light curves have power spectra that
match the hard and soft band power spectra of the simple NLS1
PG1244+026. In the subsequent panels we increase the number of
occulting clouds ($n_{cl}=10^{-4}$, $10^{-3}s^{-1}$). The occultations
are most obvious in the hard band, where the flux drops are
conspicuously narrower than in the soft band, due to the smaller
physical size of the corona compared to the more extended soft band
components. These occultations add power to the light curve. The most
heavily occulted hard band light curve (bottom right panel) shows
peaks and deep troughs more typical of a complex NLS1.

Fig. 14 shows the power spectra, lag-frequency spectra and coherence
between hard and soft bands for the same three simulations. Comparing
the hard band power spectra (Fig. 14a, dashed lines) of the original
model (black) with the most heavily obscured model (magenta) shows
that occultations have increased the power at $\sim3\times10^{-4}$Hz
by almost one and a half orders of magnitude. Hard band power spectra
of complex NLS1s routinely show similarly high power at these
frequencies.  Thus occultations are more than capable of increasing
the hard band high frequency power from $fP(f)\sim10^{-2}$ typical
of a simple NLS1 to $\sim 0.1$ as is typical of a complex NLS1. The
power increase in the soft band (solid lines) is much smaller ($\sim$
half an order of magnitude). However we have only included
occultations at $z_{cl}=0$. Higher latitude occultations would add
additional power to the soft band, however this power would be
uncorrelated with the hard band.

Fig. 14b shows the lag as a function of frequency between the hard and
soft bands. The black points show the original simple NLS1 model, with
a strong reverberation lag of $\sim200$s at $\sim5\times10^{-4}$Hz,
matching that seen in PG1244+026. As the number of occultations
increases (blue to magenta), the maximum measured reverberation lag
decreases from $200$ to $50$s and increases in frequency from
$\sim5\times10^{-4}$ to $10^{-3}$Hz. This much shorter reverberation
lag, at higher frequency, is much more typical of those seen in
complex NLS1s such as 1H0707-495. In our model, this is a direct
result of the soft leads introduced by the occultations at low
frequencies, diluting the reverberation lag and shifting its minimum
to higher frequencies. This is also in good agreement with the findings of Kara et al (2013), who showed that the reverberation lag of the complex NLS1 IRAS 13224--3809 is much shorter and at higher frequency during low flux periods (when in this scenario it would be more obscured) than high flux periods.

Fig. 14c shows the coherence between hard and soft bands. As the number
of occultations increases, the coherence drops slightly, particularly
at low frequencies (from $\sim1$ to $\sim0.8$). This is due to the
slightly different shaped flux drops in the hard and soft bands. The
sudden drop in coherence at $10^{-4}$Hz, characteristic of the simple
NLS1 PG1244+026, becomes less obvious in the more obscured
simulations. The coherence functions of complex NLS1s do show a more
gradual drop in coherence with increasing frequency, so this is not in
disagreement with the data.

\subsection{Spectral Changes}

In Fig. 12, in grey, we show an example of the total spectrum when the
source is highly absorbed. We assume the clouds are the source of the
reflected and reprocessed emission, hence these components are not
absorbed. Consequently the low flux spectrum is dominated by
reprocessed emission at low energies ($<1$keV) and reflected emission
at high energies This gives a total spectrum that is no longer
power-law like but instead shows strong curvature, with a large soft
excess and a strong iron emission line, but the 'smoking gun' signature of
highly ionised occultation is the presence of highly ionised Fe
K$\alpha$ absorption lines at 6.7 and 6.95~keV. The strength of these
features is probably underestimated in our model as the {\sc zxipcf}
model used assumes a turbulent velocity of only 200~km/s, which is
probably much smaller than expected in a failed wind structure from
the inner disc. 

Strong highly ionised absorption lines are detected in the deep dip
states of several complex NLS1 such as MRK 766 (Miller et al 2007),
MCG---6-30-15 (Miller et al 2008) and MRK 335 (Gallo et al 2013).  It
is possible that even higher turbulent velocities in the failed wind
could merge the 6.7 and 6.95~keV absorption lines into each other, and
into the absorption edge, which could produce the more dramatic drop
at K$\alpha$ in 1H0707-495 (Hagino et al, in preparation).

Our assumption that the clouds do not occult the reflected
and reprocessed emission sets a limit to the amount of absorption
present in our model. Complex NLS1s often show rather stronger drops at
low energy, which in this scenario would require that the clouds do
also occult part of the failed wind structure or that our reprocessed
emission is overestimated due to our models taking only the model flux
rather than weighting this by the instrument response.  Nonetheless,
this model demonstrates the potential of this scenario to match the
spectral variability, given a more sophisticated prescription for the
reflected/reprocessed flux.

\section{Discussion}

We have shown that occultation of the accretion flow can introduce
lags between the hard and soft energy bands. In particular the
occultations generate soft leads at a frequency related to the transit
time. For a transit time of $10^4$s, corresponding to an orbital
radius of $20R_g$, these lags predominate at low frequencies
($\sim10^{-4}$Hz for a $10^7M_{\odot}$ BH). When combined with a model
for the intrinsic variability of the accretion flow, these low
frequency soft leads act to dilute the negative reverberation lag,
reducing the maximum measured lag and shifting it to higher
frequencies, reproducing the trend seen in the data from simple to
complex NLS1. This can plausibly be produced by increasing inclination
with respect to a clumpy, turbulent structure above the inner
disc. Since NLS1s are high accretion rate sources, it is
likely that the inner regions of the disc will become 'puffed up' to a
large scale height where the local accretion rate exceeds the
Eddington limit (e.g. Jiang et al 2014), producing a turbulent
structure capable of obscuring the innermost regions. Low inclination
sources are rarely occulted (simple NLS1), while high inclination ones
have multiple occultations (complex NLS1). The assumed higher
inclination for complex NLS1s also explains the observed association
of extreme (deep dip) spectra with warm absorbers. This is not causal,
in that the warm absorbers are not distorting our view of the extreme
spectra (e.g. Chiang \& Fabian 2011), but is instead a consequence of
high inclination so that the line of sight is more likely to intercept
a wind driven from the torus/flattened broad line region (BLR).
Importantly, the occultations superimpose highly ionised absorption
lines at FeK$\alpha$ in the dips, as are seen in the data.

We show the evolution of the lag-frequency with increased occultations
for a fixed black hole mass of $10^7M_\odot$, but this should also
depend on black hole mass. The intrinsic lags/leads in the spectral
components should scale simply with mass, as should the occultation
timescale. However, mass for NLS1s is hard to determine accurately, as
they are accreting close to Eddington. Masses estimated from line
widths assume the BLR clouds are
virialised. However the effective gravity experienced by the clouds
will be reduced due to radiation pressure from the central source,
leading to an underestimate for masses of NLS1s (Marconi et al
2008). Inclination is also another uncertainty, as the BLR velocity
field is not completely virialised, but contains a clear equatorial
component (e.g. Collin et al 2006; Kollatschny \& Zetzl 2013; Pancoast
et al 2014).  Any equatorial component to the velocity field will be
suppressed in low inclination (simple) NLS1s, so their masses will be
systematically biased towards higher values compared to high inclination (complex) NLS1s.

Leighly (1999) give the FWHM for the H$\beta$ line widths in PG1244+026
and 1H0707-495 as 830 and 1050 km/s, respectively. These are corrected
for FeII and have the narrow H$\beta$ component subtracted assuming
that this is $0.1\times$ the [OIII] line intensity (see also Leighly \& Moore
2004). Both sources have intrinsic optical luminosities which are very
similar, so these give masses which are $2.4\times 10^6$ and
$3.7\times 10^6M_\odot$, respectively (Nikolajuk et al 2009). The two
object masses may be even closer if the inclination dependence
discussed above is important. An Eddington correction to the mass of
PG1244+026 increases the mass estimate to $10^7M_\odot$ (Done et al
2013), but this should also be similar for 1H0707-495. Hence, while
there are large uncertainties on masses for NLS1s, these two objects
should be very similar. At this larger mass, PG1244+026 is at the
Eddington limit for a low spin black hole (Done et al 2013). A lower
mass and/or higher spin pushes the system to higher Eddington
fractions, so making it even less likely that the disc is flat. 

Our model is more of a pilot study than a complete description.
Obvious improvements are to include general relativistic effects of
light bending on the disc image (Fig. 1a), e.g.  Middleton \& Ingram
(2014), Miniutti et al (2014). This would be most important for the
central coronal regions as the far side of this small source always
has a small impact parameter with the black hole.  This would make the
corona appear larger, so a $5R_g$ cloud may not occult the entire
corona. As a consequence the coronal flux drops would not be quite as
narrow and deep and this would reduce the difference between the total
amount of power added to the hard and soft bands. Complex NLS1s do
show more power in the hard band than the soft band (Zoghbi et al
2011) at all frequencies, unlike the simple NLS1 PG1244+026, which
shows comparable power at low frequencies in the hard and soft
bands. Our occultation model replicates this, since occultations add
power to both bands. However the small size of the corona necessarily
adds much more power to the hard band. Including light-bending (or a
smaller cloud size) would slightly lessen this difference, in better
agreement with the data.

The model presented here is in some way a composite between the
previous extreme relativistic reflection models and partial covering
models. It follows the partial covering model in identifying
absorption (as opposed to light bending) events as the origin for the
deep dips, but has the occulting material be closer to the source 
(ten gravitational radii rather than a few tens-hundreds), and be more
highly ionised. Reflection does make an important contribution to the
spectrum during the dips in our model, but it is not extremely smeared
by relativistic effects. Instead, and in a step beyond what is modeled
here, we envisage the reflector as a clumpy,
turbulent, failed wind rather than a flat Keplerian disc (see also
Miller et al 2008). The clumps may be only marginally optically thick,
so their reflected/scattered emission is not quite the same shape as
from $\tau\gg 1$ (Miller \& Turner 2013), and they may be embedded in
hotter material which Comptonises the reflected
emission. Additionally, the cloud itself could have complex structure
due to the ionisation instability of X-ray illuminated material
(Krolik, McKee \& Tarter 1981). The illuminated face of the cloud will
be heated to the local Compton temperature ($\sim 10^6$K). Temperature
decreases at larger depths into the cloud where scattering reduces the
heating, so the density must increase to keep in pressure
balance. This lower ionisation state material has more line cooling,
so the temperature drops abruptly, giving a sharp transition between a
highly ionised skin and a nearly neutral core (Chevalier et al
2006). Reflection from such structures, especially with a turbulent
velocity field, may be a feasible way to reproduce the observed 2-10~keV
spectra in the dips.

Another difference between this model and standard relativistic
reflection is that we include thermalisation of the illuminating
flux. Hard X-rays which are not reflected can either heat the disc and
be re-emitted as (mostly) thermalised radiation, or the energy can be
released as lines/recombination continua.  The relative importance of
these two processes depends strongly on the vertical structure of the
disc. Thermalisation is more important for discs in hydrostatic
equilibrium (Nayakshin et al 2001) but current reflection models are
calculated for constant density discs (e.g. Ross \& Fabian 2005;
Garcia et al 2014).  The high disc temperatures expected in NLS1s means
that this component must be important at some level in contributing to
the soft X-ray excess in these objects, and since it is predominantly
thermal then it has no strong soft X-rays lines which require high
spin to smear them into the observed smooth continuum.

However, probably the most important effect which should be included
in matching to real data is that the light curves in the soft and hard
bands are weighted by the detector response rather than being simply
flux integrated over the energy band as used here. This is a key
requirement to fit the model to real data, which will be the subject
of  a subsequent paper. 

\section{Conclusions}

We have constructed a simple occultation model to investigate whether
the change in spectral and timing properties between simple and
complex NLS1s can be explained by a difference in inclination with
respect to a failed wind. In this scenario, clumps of material lifted
from the inner parts of the accretion disc by radiation pressure,
obscure the X-ray emission for sources seen at high inclination
angles, resulting in more extreme variability and more complex
spectra. Associating the deep dips with occultation superimposes
absorption features from FeK$\alpha$ on the dip spectra. This is seen
in complex NLS1s such as MRK 776 (Miller et al 2007) and is a smoking
gun for the reality of these occultation events.

We model the obscuration as a series of individual clouds of constant
ionisation parameter which transit the inner accretion flow,
co-rotating with the flow and obscure the underlying emission. The underlying accretion flow emission is radially stratified, with the softest X-rays (disc) from the largest
radii, and then the soft X-ray excess and corona at progressively
smaller radii. We find that occultations add power to the X-ray light
curves over a range of frequencies related to the transit
time. Occultations also introduce a lag between the hard and soft
energy bands when Doppler boosting of the underlying accretion flow
emission is taken into account; specifically occultations introduce a
soft lead, with the hard band lagging the soft band.

We then combined our occultation model with the full spectral-timing
model of GD14 which describes the accretion flow emission of the
simple NLS1 PG1244+026. This model also includes reprocessed emission
as part of the soft X-ray excess, as well as reflection from it. 
This reproduces the timing properties of PG1244+026 by
assuming slow fluctuations are generated in the outer components and
these propagate down to the corona, producing low frequency hard
lags. The high frequency soft lags (reverberation lags) are produced
predominantly by fast coronal fluctuations being reprocessed in the
soft excess wind material rather than by reflection from it.

The effect of the occultations is to dilute the negative reverberation
lag and shift it to higher frequencies. By increasing the rate of
occultations we can match the change in reverberation lag from the
$200$s at $5\times10^{-4}$Hz seen in the simple NLS1 PG1244+026, to
the $\sim50$s lag at $10^{-3}$Hz seen in the complex NLS1
1H0707-495. The lag times and light travel times put into the model
are the same in both the obscured and unobscured cases. The only
difference is the presence of occultations. It is the soft leads
caused by the broken symmetry of the ingress and egress of the
occultations caused by Doppler boosting of the disc emission which
result in a shorter net reverberation lag in the obscured case.  The
occultations also change the energy spectrum from a simple NLS1, with
strong soft excess and a steep power law above 2 keV, to
something resembling a complex NLS1, with a prominent iron line from
reflection off the soft excess, highly ionised Fe K$\alpha$ absorption lines and strong spectral curvature. 

The short $\sim50$s reverberation lags have been taken as evidence for
extreme relativistic reflection in complex NLS1s. If 50s really is a
light travel time this requires reflection from the innermost radii of
an accretion disc around a highly spinning BH. Our occultation model
shows that this need not be the case. A short reverberation lag can be
the result of a much longer light travel time, diluted and shifted by
the soft leads introduced by occulting clouds. In our model,
reflection and reprocessing occurs between $6-12R_g$ and puts no
constraints on the spin of the BH. Changing inclination then naturally
explains the change from smooth spectra to complex spectra, and long
lower frequency reverberation lags to shorter higher frequency
reverberation lags in simple and complex NLS1s. Given that NLS1s are
high accretion rate sources, it is quite natural to expect that the
disc is not flat, that radiation pressure can lift material from
the disc which will obscure the central emission for high inclination
lines of sight. This is a promising geometry to explore further. 

\section{Acknowledgements}

We thank the anonymous referee for helpful comments. EG acknowledges useful discussions with Will Alston and Matt Middleton. CD acknowledges useful conversations with Giovanni Miniutti. EG acknowledges funding from the UK STFC.

\label{lastpage}

\end{document}